\documentclass[
aps,
twocolumn,
longbibliography,
nofootinbib
]{revtex4-2}
\usepackage[T1]{fontenc}% needed for characters such as \k in the bibliography
\usepackage{amsthm}
\theoremstyle{plain}
\newtheorem{theorem}{Theorem}%[section]
\newtheorem*{theorem*}{Theorem}

\newtheorem*{definition*}{Definition}

\newtheorem*{lemma*}{Lemma}
\newtheorem{corollary}[theorem]{Corollary}

\usepackage{braket}
\usepackage{delarray}
\usepackage{here}

\usepackage{amsmath}
\usepackage{txfonts}

\usepackage{iftex}
\ifpdf
  \usepackage{graphicx}
\else
  \usepackage[dvipdfmx]{graphicx}
\fi
\usepackage{bm}
\usepackage{color}
\definecolor{revblue}{rgb}{0,0,0.6}% color for revised text
\usepackage[normalem]{ulem}% normalem: keep \emph in italics (needed for book titles in the bibliography)

\usepackage{physics}

\ifpdf
  \usepackage[colorlinks=true,linkcolor=blue,citecolor=blue,urlcolor=blue]{hyperref}
\else
  \usepackage[dvipdfmx,colorlinks=true,linkcolor=blue,citecolor=blue,urlcolor=blue]{hyperref}
\fi

\newcommand{\ba}{\begin{array}}
\newcommand{\ea}{\end{array}}
\newcommand{\bmat}{\left(\begin{array}}
\newcommand{\emat}{\end{array}\right)}

\newcommand{\be}{\begin{eqnarray}}
\newcommand{\ee}{\end{eqnarray}}

\newcommand{\E}{\mathbb{E}}

\newcommand{\bsig}{\bm{\sigma}}
\newcommand{\bet}{\bm{\eta}}
\newcommand{\bu}{\bm{u}}
\newcommand{\heta}{\hat{\eta}}
\newcommand{\tet}{\tilde{\bm{\eta}}}

\begin{document}
\title{Spin-glass order in a field requires magnetic-field chaos in XY gauge glasses for $d\leq 4$}
\author{Manaka Okuyama$^{1,2}$}
\thanks{Corresponding author: manaka.okuyama.d2@tohoku.ac.jp}
\author{Masayuki Ohzeki$^{1,3,4,5}$}
\author{Hidetoshi Nishimori$^{1,6}$}
\affiliation{$^1$Graduate School of Information Sciences, Tohoku University, Sendai 980-8579, Japan}
\affiliation{$^2$School of Computing, Institute of Science Tokyo, Tokyo 152-8551, Japan}
\affiliation{$^3$Department of Physics, Institute of Science Tokyo, Tokyo 152-8551, Japan}
\affiliation{$^4$Research and Education Institute for Semiconductors and Informatics, Kumamoto University, Kumamoto 860-0862, Japan}
\affiliation{$^5$Sigma-i Co., Ltd., Tokyo 108-0075, Japan} %\\
\affiliation{$^6$Institute of Integrated Research, Institute of Science Tokyo, Nagatsuta-cho, Midori-ku, Yokohama 226-8501, Japan}

\begin{abstract}
The XY gauge-glass model is believed to exhibit a spin-glass phase in dimensions $d\geq 3$. In this paper, we prove that, in dimensions $d\leq 4$ and at any finite temperature, a magnetic field applied in a fixed direction, with amplitudes that fluctuate from site to site according to the Rayleigh distribution, must either induce magnetic-field chaos in the spin-glass phase or destroy spin-glass order altogether, whenever a spin-glass phase exists at zero field. In particular, if a spin-glass phase persists under such a field, as predicted by the replica-symmetry-breaking (RSB) picture, it must exhibit magnetic-field chaos.
This conclusion follows from our proof that, for any fixed nonzero field strength, the disorder average of the squared overlap between equilibrium spin configurations at zero field and those at nonzero field vanishes in the thermodynamic limit. By gauge transformations, the overlap statistics are identical to the magnetization statistics of an XY gauge glass with correlated bond disorder in a Gaussian random field. Our proof combines this exact identity with quantitative estimates on the suppression of ferromagnetic order by random fields in continuous-spin systems, due to Dario, Harel, and Peled [\href{https://doi.org/10.1007/s00220-024-05081-9}{Commun.\ Math.\ Phys.\ \textbf{405}, 212 (2024)}] and building on the work of Imry and Ma [\href{https://doi.org/10.1103/PhysRevLett.35.1399}{Phys.\ Rev.\ Lett.\ \textbf{35}, 1399 (1975)}] and of Aizenman and Wehr [\href{https://doi.org/10.1103/PhysRevLett.62.2503}{Phys.\ Rev.\ Lett.\ \textbf{62}, 2503 (1989)}; \href{https://doi.org/10.1007/BF02096933}{Commun.\ Math.\ Phys.\ \textbf{130}, 489 (1990)}]. We extend these estimates to bond-dependent interactions. Although our result provides a rare rigorous constraint on the nature of spin-glass phases in finite-dimensional systems, it does not determine whether spin-glass order survives in a field, and therefore does not distinguish between the droplet and RSB pictures. The case of a strictly uniform field remains open.
\end{abstract}
\date{\today}
\maketitle

%%%%%%%%%%%%%%%%%%%%%%%%%%%%%%%%%%%%%%%%%%%%%%%%%%%%%%%%%%%%%%%%%%%%%%%%%%%
%%%%%%%%%%%%%%%%%%%%%%%%%%%%%%%%%%%%%%%%%%%%%%%%%%%%%%%%%%%%%%%%%%%%%%%%%%%
\section{Introduction}

The development of interpolation methods has led to remarkable progress in the rigorous analysis of mean-field spin-glass models~\cite{GuerraToninelli2002,Guerra2003,Talagrand2006,Panchenko2013}, such as the Sherrington--Kirkpatrick model~\cite{Sherrington1975}, whose solution had been obtained by the replica method~\cite{Parisi1979,Parisi1980a,Mezard1987,Charbonneau2023}.
In contrast, the rigorous analysis of finite-dimensional spin-glass models~\cite{Edwards1975} remains a formidable challenge, with only limited results available. In particular, whether replica symmetry breaking (RSB), as predicted by mean-field theory, persists in finite dimensions remains a subject of longstanding debate~\cite{Fisher1986,Fisher1988,SteinNewman2013,Charbonneau2023}.

Gauge symmetry is one of the few analytical tools that yield rigorous results for finite-dimensional spin glasses, including constraints on phase diagrams~\cite{Nishimori1981,Horiguchi1982,Nishimori1986,Kitatani1992} and an exact expression for the internal energy on the Nishimori line~\cite{Nishimori1980,Nishimori1981,Nishimori2001}. For continuous spins, gauge symmetry has also been used to prove the breaking of continuous symmetry on the Nishimori line in three dimensions~\cite{Garban2022}. Nevertheless, obtaining rigorous information about the properties of the spin-glass phase itself has long remained difficult.

Recently, however, progress has been made in understanding the properties of spin-glass phases by exploiting gauge symmetry. In particular, a previous study~\cite{Nishimori2025a}
has established a nontrivial relation between temperature chaos~\cite{Bray1987,Banavar1987,Kondor1989,Rizzo2003,Katzgraber2007b,Fernandez2013,BaityJesi2021,Chen2013,Panchenko2016} and reentrant phase transitions~\cite{Nobre2001,Hasenbusch2007,Ceccarelli2011} in the Ising spin-glass models in arbitrary dimensions by exploiting gauge symmetry in spin-glass models with a specific form of correlated disorder~\cite{Kitatani1992,Nishimori2022,Nishimori2024,Nishimori2025}; spin glasses with other types of correlated disorder have also been studied~\cite{Hoyos2011,Bonzom2013,Cavaliere2019,Munster2021}. Specifically, the absence of temperature chaos in the spin-glass phase implies that the phase boundary between the spin-glass and ferromagnetic phases must be vertical in the corresponding phase diagram. Remarkably, this relation follows from general properties of gauge symmetry and extends to a broad class of spin-glass models, including XY gauge glasses~\cite{Ozeki1993} and gauge-invariant Potts spin glasses~\cite{Nishimori1983}. An explicit realization of this relation has been found in the fully connected Ising $p$-spin glass~\cite{Nishimori2026pspin}.

In this paper, we focus on the XY gauge-glass model, introduced as a model of vortex glasses in disordered superconductors~\cite{FisherMPA1989,HuseSeung1990}, which is believed to exhibit a spin-glass phase in three or more dimensions~\cite{Reger1991,Ozeki1993,KosterlitzAkino1998,Olson2000,Katzgraber2005,Alba2011}.
By further developing an analytical framework that combines gauge symmetry with spin-glass models with correlated disorder, we establish rigorous constraints on the response of the spin-glass phase to a magnetic field in three and four dimensions. The field is applied in a fixed direction, and its amplitude fluctuates from site to site around a nonzero mean value (Sec.~\ref{sec:model}).

The stability of spin-glass phases under a magnetic field has long been a central problem in spin-glass physics. According to the mean-field RSB picture, (i) the spin-glass phase remains stable under a sufficiently weak magnetic field~\cite{deAlmeida1978,Parisi1979}, which nevertheless induces a drastic rearrangement of equilibrium spin configurations; this phenomenon is known as magnetic-field chaos~\cite{Kondor1989,Ritort1994,Billoire2003,AguilarJanita2024}.
In contrast, the droplet picture predicts that (ii) an arbitrarily weak magnetic field destroys the spin-glass phase~\cite{Fisher1986,Bray1987,Fisher1988}. Whether a spin-glass phase survives in a magnetic field in finite dimensions remains controversial, mainly in studies of Ising spin glasses~\cite{YoungKatzgraber2004,MooreBray2011,Janus2012}.
A third scenario is also logically possible: (iii) the spin-glass phase persists under a magnetic field without exhibiting magnetic-field chaos. We rigorously prove that this third scenario is impossible for the XY gauge-glass model in three and four dimensions under such a field. In other words, if the spin-glass phase survives in a weak field, as predicted by the RSB picture, magnetic-field chaos necessarily follows. To our knowledge, this implication has not been established in finite dimensions, even under the assumption that the spin-glass phase survives in a field.
Although our result does not distinguish between the RSB and droplet pictures, it provides one of the few rigorous constraints on the nature of spin-glass phases in finite dimensions.

The key to our proof is an exact identity derived from gauge symmetry that relates the spin overlap between equilibrium configurations at zero and nonzero magnetic fields in the XY gauge-glass model to the magnetization of an XY gauge-glass model with correlated disorder in a Gaussian random field. This identity converts the question of how the spin-glass phase responds to a magnetic field into the question of whether a random field destroys ferromagnetic order, a question that has been studied extensively for random-field systems.
An identity of the same form was previously established for the Ising spin-glass model~\cite{Nishimori2025}.

We then show that the correlated model in a random field has no ferromagnetic long-range order in dimensions $d\leq 4$. This step relies on the Imry--Ma mechanism~\cite{Imry1975},
established rigorously by Aizenman and Wehr~\cite{AizenmanWehr1989,AizenmanWehr1990}
and made quantitative by Dario, Harel, and Peled~\cite{Dario2024}:
the energy cost of a spin wave that reorients the spins in a region of linear size $\ell$, which is of order $\ell^{d-2}$, is overwhelmed by the random-field energy that can be gained, which is of order $h\ell^{d/2}$, for $d<4$, while $d=4$ is the marginal case.
The arguments of Dario, Harel, and Peled are formulated for bond-independent interactions.
We extend their approach to allow for bond-dependent interactions and apply it to the correlated model. The exact identity~\cite{Nishimori2025} then implies that the spin overlap between equilibrium configurations at zero and nonzero magnetic fields in the XY gauge-glass model vanishes in the thermodynamic limit.

Consequently, our result rigorously rules out the possibility that a spin-glass phase persists under a magnetic field without exhibiting magnetic-field chaos in the XY gauge-glass model in three and four dimensions. Since the Imry--Ma mechanism does not depend on the nature of the bond disorder, our result may be viewed as a translation, by means of gauge symmetry, of a general property of low-dimensional continuous-spin systems into a constraint on the spin-glass phase.

The rest of this paper is organized as follows. In Sec.~\ref{sec:model}, we introduce the models, derive the exact identities, and state the main results. Section~\ref{sec:discussion} discusses the results and their limitations. The use of generative-AI tools is described in the \hyperref[sec:methods]{Methods section}. The proof of Theorem~\ref{thm:main} is given in Appendix~\ref{app:proof}.

%%%%%%%%%%%%%%%%%%%%%%%%%%%%%%%%%%%%%%%%%%%%%%%%%%%%%%%%%%%%%%%
\section{Model and main result}\label{sec:model}

We consider the XY gauge-glass model in a Gaussian
random field.
Let $\Lambda_L$ be the $d$-dimensional hypercubic
lattice of linear size $L$ with periodic boundary conditions, with $N=|\Lambda_L|=L^d$ sites and $N_b=dN$ bonds. Free boundary conditions can be treated in the same way. With fixed boundary conditions, the gauge transformations used below would also change the boundary spins, and the identities derived below would have to be modified.
The Hamiltonian is given by
\begin{equation}
H_{\mathrm{RF}}(h)
=
-\sum_{\langle ij\rangle}
\cos(\theta_i-\theta_j-A_{ij})
-h\sum_i
\left(
\eta_i^x\cos\theta_i
+\eta_i^y\sin\theta_i
\right),
\label{eq:random_field}
\end{equation}
where $\theta_i\in[0,2\pi)$ denotes the angle of the
XY spin at site $i$, and $h\geq0$ controls the
strength of the random field.
The random variables $\eta_i^x$ and $\eta_i^y$
are independent standard Gaussian random variables,
while the bond variables $A_{ij}$ are independently
and uniformly distributed over $[0,2\pi)$,
with $A_{ji}=-A_{ij}$ modulo $2\pi$.
All random variables are mutually independent,
apart from this antisymmetry constraint.

Introducing the polar representation
\begin{equation}
\eta_i^x=r_i\cos\phi_i,
\qquad
\eta_i^y=r_i\sin\phi_i,
\end{equation}
we can rewrite the Hamiltonian as
\begin{equation}
H_{\mathrm{RF}}(h)
=
-\sum_{\langle ij\rangle}
\cos(\theta_i-\theta_j-A_{ij})
-h\sum_i r_i\cos(\theta_i-\phi_i).
\label{eq:random_field_polar}
\end{equation}
Here, the random angles $\phi_i$ are independently
and uniformly distributed over $[0,2\pi)$,
while the amplitudes $r_i$ are independently
distributed according to the Rayleigh distribution
with probability density
\begin{equation}
p(r)=r e^{-r^2/2},
\qquad r\geq0.
\end{equation}
The random angles and amplitudes are mutually
independent and are also independent of the
bond variables.
For later use, we introduce the normalized integration measures
\begin{align}
d\mu(A)
&:=\prod_{\langle ij\rangle}\frac{dA_{ij}}{2\pi},
\qquad
d\nu(\theta)
:=\prod_i\frac{d\theta_i}{2\pi},
\nonumber\\
d\rho(r)
&:=\prod_i r_i e^{-r_i^2/2}\,dr_i,
\label{eq:measures}
\end{align}
and denote the average over the Gaussian random field by
\begin{align}
\E_\eta[X]
&:=\int\prod_i\frac{d\eta_i^x\,d\eta_i^y}{2\pi}\,
e^{-[(\eta_i^x)^2+(\eta_i^y)^2]/2}\,X
\nonumber\\
&\,=\int d\rho(r)\,d\nu(\phi)\,X .
\label{eq:Eeta}
\end{align}

We next consider the Hamiltonian obtained by setting
$\phi_i=0$ at every site:
\begin{equation}
H_{\mathrm{F}}(h)
=
-\sum_{\langle ij\rangle}
\cos(\theta_i-\theta_j-A_{ij})
-h\sum_i r_i\cos\theta_i.
\label{eq:biased_field}
\end{equation}
This model describes an XY gauge glass subjected
to a spatially varying magnetic field whose
magnitude differs from site to site, while its
direction is fixed at $\theta=0$.
The random amplitudes $r_i\geq0$ remain independently
distributed according to the Rayleigh distribution.

The mean of the Rayleigh distribution is $\sqrt{\pi/2}$. Writing $r_i=\sqrt{\pi/2}+\delta r_i$, where $\delta r_i$ has zero mean, the field term in Eq.~\eqref{eq:biased_field} is the sum of a uniform field of strength $h\sqrt{\pi/2}$ in the direction $\theta=0$ and a random field $h\,\delta r_i$ along the same axis, which has zero mean, takes both signs, and fluctuates independently from site to site. Equation~\eqref{eq:biased_field} thus describes the XY gauge glass under a magnetic field applied in a fixed direction, whose magnitude fluctuates from site to site. The relative fluctuation of the amplitude, $\sqrt{\mathrm{Var}(r_i)}/\E[r_i]=\sqrt{(4-\pi)/\pi}\simeq0.52$, does not depend on $h$; the field is therefore not a weak perturbation of a uniform field, even for small $h$. The case of a strictly uniform field is discussed in Sec.~\ref{sec:discussion}. In the context of vortex glasses, we note that $h$ is a field conjugate to the XY spins; it does not correspond to the physical magnetic field applied to a superconductor, which acts through the vector potential.

Finally, we consider the XY gauge-glass model with
correlated bond disorder in a Gaussian random field.
The Hamiltonian is given by
\begin{equation}
H_{\mathrm{corr}}(h)
=
-\sum_{\langle ij\rangle}
\cos(\theta_i-\theta_j-B_{ij})
-h\sum_i r_i\cos(\theta_i-\phi_i).
\label{eq:correlated_model}
\end{equation}
The bond variables $B_{ij}$ are distributed according
to the joint probability density
\begin{equation}
P_{\mathrm{corr}}(B)
=
\frac{1}{(2\pi)^{N_b}}
\frac{
\exp\left[
\beta\sum_{\langle ij\rangle}\cos B_{ij}
\right]
}{
Z_0(B;\beta)
},
\label{eq:Pcorr}
\end{equation}
where $\beta$ is the inverse temperature and
$Z_0(B;\beta)$ denotes the partition function
of the XY gauge-glass model in the absence
of a magnetic field:
\begin{equation}
Z_0(B;\beta)
=
\int
d\nu(\theta)\,
\exp\left[
\beta\sum_{\langle ij\rangle}
\cos(\theta_i-\theta_j-B_{ij})
\right].
\end{equation}
The probability density is defined with respect
to the product measure
$\prod_{\langle ij\rangle}dB_{ij}$,
with $B_{ji}=-B_{ij}$ modulo $2\pi$.
The normalization factor $(2\pi)^{-N_b}$ is exact.
This follows from the change of variables $B_{ij}\to B_{ij}+\chi_i-\chi_j$ with arbitrary site angles $\chi_i$, which leaves $Z_0(B;\beta)$ invariant, followed by the average over $\chi$:
\begin{align}
&\int\prod_{\langle ij\rangle}dB_{ij}\,
\frac{e^{\beta\sum_{\langle ij\rangle}\cos B_{ij}}}{Z_0(B;\beta)}
\nonumber\\
&=(2\pi)^{N_b}\!\int d\mu(B)\,d\nu(\chi)\,
\frac{e^{\beta\sum_{\langle ij\rangle}\cos(B_{ij}+\chi_i-\chi_j)}}{Z_0(B;\beta)}
=(2\pi)^{N_b},
\label{eq:Pcorr_normalization}
\end{align}
since the integral over $\chi$ reproduces $Z_0(B;\beta)$.
Unlike the conventional XY gauge-glass model,
the bond variables are generally correlated
through the partition function $Z_0(B;\beta)$.
The Gaussian random-field components
$\eta_i^x$ and $\eta_i^y$ remain independent
standard Gaussian random variables and are
independent of the bond disorder.

The distribution~\eqref{eq:Pcorr} is defined separately for each system size $L$. No consistency between the distributions for different sizes is required below, because the estimates we use are uniform in the bond configuration.
The parameter $\beta$ appearing in the
disorder distribution can generally differ from
the inverse temperature of the system~\cite{Nishimori2024,Nishimori2025}.
We first choose them to coincide and discuss the general case in Sec.~\ref{sec:twoT}.

For each model
$\mathsf M\in\{\mathrm F,\mathrm{RF},\mathrm{corr}\}$,
we denote the thermal expectation at field strength
$h$ by
\begin{equation}
\langle O\rangle_{\mathsf M,h}
:=\frac{1}{Z_{\mathsf M}(h)}\int d\nu(\theta)\,O(\theta)\,e^{-\beta H_{\mathsf M}(h)},
\end{equation}
where $Z_{\mathsf M}(h)=\int d\nu(\theta)\,e^{-\beta H_{\mathsf M}(h)}$ depends on the realization of the disorder.
The joint thermal expectation over two independent
equilibrium configurations $\theta^{(1)}$ and $\theta^{(2)}$ at field strengths
$h$ and $0$, respectively, sampled with the same realization
of the disorder, is denoted by
$\langle\cdot\rangle_{\mathsf M;h,0}$.
It is defined by the product
of the corresponding Gibbs distributions:
\begin{equation}
\langle\cdot\rangle_{\mathsf M;h,0}
=
\langle\cdot\rangle_{\mathsf M,h}
\otimes
\langle\cdot\rangle_{\mathsf M,0}.
\label{thermal-average}
\end{equation}
For any disorder-dependent physical quantity $X$, we define its disorder averages in the three models by
\begin{equation}
\begin{aligned}
\mathbb E_{\mathrm F}[X]
&:=\int d\mu(A)\,d\rho(r)\,X,\\
\mathbb E_{\mathrm{RF}}[X]
&:=\int d\mu(A)\,\E_\eta[X]
=\int d\mu(A)\,d\rho(r)\,d\nu(\phi)\,X,\\
\mathbb E_{\mathrm{corr}}[X]
&:=\int d\mu(B)\,
\frac{e^{\beta\sum_{\langle ij\rangle}\cos B_{ij}}}{Z_0(B;\beta)}\,
\E_\eta[X].
\end{aligned}
\label{eq:disorder_averages}
\end{equation}
In the last expression, the normalization factor $(2\pi)^{-N_b}$ is absorbed into $d\mu(B)$. Note that the random field is independent of the bonds under $\mathbb E_{\mathrm{corr}}$; this fact plays an essential role below.

We define the complex spin overlap between two spin configurations $\theta^{(1)}$ and $\theta^{(2)}$, and the complex magnetization, by
\begin{equation}
Q_{\Lambda_L}
:=
\frac{1}{N}\sum_{i\in\Lambda_L}
e^{i(\theta_i^{(1)}-\theta_i^{(2)})},
\qquad
M_{\Lambda_L}
:=
\frac{1}{N}\sum_{i\in\Lambda_L}
e^{i\theta_i}.
\label{eq:QM}
\end{equation}
Our primary interest is the distribution function of the spin overlap between equilibrium configurations at zero and nonzero magnetic fields in the XY gauge-glass model~\eqref{eq:biased_field}. For a system of size $L$, it is
\begin{equation}
P_{\mathrm F}^{(L)}(q;0,h)
:=
\mathbb E_{\mathrm F}
\left[
\left\langle
\delta^{(2)}(q-Q_{\Lambda_L})
\right\rangle_{\mathrm F;h,0}
\right],
\end{equation}
where $q\in\mathbb C$ and $\delta^{(2)}(q)=\delta(\mathrm{Re}\,q)\,\delta(\mathrm{Im}\,q)$ denotes the two-dimensional Dirac delta function. According to the definition of the thermal average in Eq.~\eqref{thermal-average}, the first replica $\theta^{(1)}$ corresponds to the Hamiltonian~\eqref{eq:biased_field} at field strength $h$, while the second replica $\theta^{(2)}$ corresponds to the same Hamiltonian at $h=0$.
Similarly, we define
\begin{align}
P_{\mathrm{RF}}^{(L)}(q;0,h)
&:=
\mathbb E_{\mathrm{RF}}
\left[
\left\langle
\delta^{(2)}(q-Q_{\Lambda_L})
\right\rangle_{\mathrm{RF};h,0}
\right],
\\
P_{\mathrm{corr}}^{(L)}(m;h)
&:=
\mathbb E_{\mathrm{corr}}
\left[
\left\langle
\delta^{(2)}(m-M_{\Lambda_L})
\right\rangle_{\mathrm{corr},h}
\right].
\end{align}
In $P_{\mathrm{RF}}^{(L)}$, the applied field is a Gaussian random field with strength parameter $h$.
We write $P_{\mathrm F}(q;0,h)$, $P_{\mathrm{RF}}(q;0,h)$, and $P_{\mathrm{corr}}(m;h)$ for the limits $L\to\infty$ when they exist. We do not assume their existence in advance; statements of the form $P_{\mathrm F}(q;0,h)=\delta^{(2)}(q)$ below are understood in the sense explained in the next paragraph.
As discussed below, $P_{\mathrm F}(q;0,h)$ provides a characterization of magnetic-field chaos.

Since $|Q_{\Lambda_L}|\leq 1$, the statement $P_{\mathrm F}(q;0,h)=\delta^{(2)}(q)$, in the sense that $\int d^2q\,\psi(q)P^{(L)}_{\mathrm F}(q;0,h)\to\psi(0)$ for every continuous function $\psi$, is equivalent to
\begin{equation}
\lim_{L\to\infty}
\mathbb E_{\mathrm F}\left[\left\langle|Q_{\Lambda_L}|^2\right\rangle_{\mathrm F;h,0}\right]=0 .
\label{eq:second_moment}
\end{equation}
Indeed, $\psi(q)=|q|^2$ is continuous on the unit disk. Conversely, if Eq.~\eqref{eq:second_moment} holds, the probability that $|Q_{\Lambda_L}|>\epsilon$ is bounded by $\epsilon^{-2}\mathbb E_{\mathrm F}[\langle|Q_{\Lambda_L}|^2\rangle_{\mathrm F;h,0}]\to0$ for every $\epsilon>0$, and the convergence follows. We use the form~\eqref{eq:second_moment} below. It does not assume the existence of the limiting distribution in advance; the convergence to $\delta^{(2)}(q)$ follows from it. The same applies to $P_{\mathrm{RF}}$ and $P_{\mathrm{corr}}$.

We also note a symmetry of $P^{(L)}_{\mathrm F}$. The zero-field replica is invariant under the global rotation $\theta_i^{(2)}\to\theta_i^{(2)}+\alpha$, which changes $Q_{\Lambda_L}$ into $Q_{\Lambda_L}e^{-i\alpha}$. Hence $P^{(L)}_{\mathrm F}(q;0,h)$ is invariant under $q\to qe^{i\alpha}$, and $\mathbb E_{\mathrm F}[\langle Q_{\Lambda_L}\rangle_{\mathrm F;h,0}]=0$ holds trivially for every $L$. The nontrivial content of Eq.~\eqref{eq:second_moment} is the concentration of $|Q_{\Lambda_L}|$ at zero.

%%%%%%%%%%%%%%%%%%%%%%%%%%%%%%%%%%%%%%%%%%%%%%%%%%%%%%%%%%%%%%%
\subsection{Exact identities from gauge transformations}\label{sec:identities}

An important consequence of gauge symmetry is that the three distribution functions introduced above coincide exactly for every system size $L$, as in the analogous identity previously established for the Ising spin-glass model~\cite{Nishimori2025}:
\begin{equation}
P_{\mathrm F}^{(L)}(q;0,h)
=
P_{\mathrm{RF}}^{(L)}(q;0,h)
=
P_{\mathrm{corr}}^{(L)}(q;h).
\label{eq:main_identity}
\end{equation}
Equivalently, for every bounded measurable function $\psi$ on the unit disk,
\begin{align}
\mathbb E_{\mathrm F}\left[\langle\psi(Q_{\Lambda_L})\rangle_{\mathrm F;h,0}\right]
&=\mathbb E_{\mathrm{RF}}\left[\langle\psi(Q_{\Lambda_L})\rangle_{\mathrm{RF};h,0}\right]
\nonumber\\
&=\mathbb E_{\mathrm{corr}}\left[\langle\psi(M_{\Lambda_L})\rangle_{\mathrm{corr},h}\right].
\label{eq:identity_psi}
\end{align}
The same equalities then hold for the limiting distributions whenever they exist.

We first establish the first equality.
In the two-replica average of the random-field model~\eqref{eq:random_field_polar}, we fix $(A,r,\phi)$ and change the integration variables as
\begin{equation}
\theta_i^{(a)}=\theta_i'^{(a)}+\phi_i\quad(a=1,2),
\qquad
A_{ij}=A'_{ij}+\phi_i-\phi_j .
\label{eq:gauge_transformation}
\end{equation}
In the usual terminology, this is the gauge transformation $\theta_i\to\theta_i-\phi_i$, $A_{ij}\to A_{ij}-\phi_i+\phi_j$; we write it as an explicit change of variables to fix its direction.
Then
\begin{align}
\cos(\theta_i-\theta_j-A_{ij})&=\cos(\theta'_i-\theta'_j-A'_{ij}),
\nonumber\\
r_i\cos(\theta_i-\phi_i)&=r_i\cos\theta'_i,
\nonumber\\
\theta_i^{(1)}-\theta_i^{(2)}&=\theta_i'^{(1)}-\theta_i'^{(2)},
\end{align}
so that the Hamiltonian~\eqref{eq:random_field_polar} with $(\theta,A,r,\phi)$ equals the fixed-direction field Hamiltonian~\eqref{eq:biased_field} with $(\theta',A',r)$, both at field $h$ and at field $0$, and the complex spin overlap remains unchanged.
Consequently, $Z_{\mathrm{RF}}(h)$ with $(A,r,\phi)$ equals $Z_{\mathrm F}(h)$ with $(A',r)$, and $\langle\psi(Q_{\Lambda_L})\rangle_{\mathrm{RF};h,0}$ with $(A,r,\phi)$ equals $\langle\psi(Q_{\Lambda_L})\rangle_{\mathrm F;h,0}$ with $(A',r)$.
Furthermore, since the bond variables are independently and uniformly distributed, $d\mu(A)=d\mu(A')$, and the transformed bond variables remain independent of the amplitudes $r_i$.
The integrand no longer depends on $\phi$, and the average over $\phi$ is trivial. We therefore obtain
\begin{align}
&
\mathbb E_{\mathrm F}
\left[
\left\langle
\psi(Q_{\Lambda_L})
\right\rangle_{\mathrm F;h,0}
\right]
=
\mathbb E_{\mathrm{RF}}
\left[
\left\langle
\psi(Q_{\Lambda_L})
\right\rangle_{\mathrm{RF};h,0}
\right]
\end{align}
for every finite system size.

We next prove the second equality in
Eq.~\eqref{eq:identity_psi}.
In the correlated XY gauge-glass model~\eqref{eq:correlated_model}, the average of $\psi(M_{\Lambda_L})$ can be written as
\begin{align}
&
\mathbb E_{\mathrm{corr}}
\left[
\left\langle
\psi(M_{\Lambda_L})
\right\rangle_{\mathrm{corr},h}
\right]
\nonumber\\
={}&
\int d\mu(B)\,d\rho(r)\,d\nu(\phi)\,
\frac{
\exp\left[
\beta\sum_{\langle ij\rangle}\cos B_{ij}
\right]
}{Z_0(B;\beta)}
\nonumber\\
&\times
\frac{
\displaystyle
\int d\nu(\theta)\,
\psi
\left(
\frac{1}{N}\sum_i e^{i\theta_i}
\right)
e^{-\beta H_{\mathrm{corr}}(h)}
}{
Z_h(B,r,\phi)
},
\label{eq:correlated_distribution_integral}
\end{align}
where $Z_h(B,r,\phi)$ denotes the partition
function of the correlated model at field strength
$h$, and the Hamiltonian in the integrand is
evaluated at $(\theta,B,r,\phi)$.
For arbitrary fixed site angles $\chi_i$, we change the integration variables as
\begin{equation}
\theta_i=\theta'_i+\chi_i,
\quad
\phi_i=\phi'_i+\chi_i,
\quad
B_{ij}=B'_{ij}+\chi_i-\chi_j .
\label{eq:gauge_transformation2}
\end{equation}
Under this change of variables, $H_{\mathrm{corr}}(h)$ with $(\theta,B,r,\phi)$ equals $H_{\mathrm{corr}}(h)$ with $(\theta',B',r,\phi')$, the partition functions $Z_h$ and $Z_0$ are invariant, and the integration measures are invariant. Only two quantities change: $\cos B_{ij}=\cos(B'_{ij}+\chi_i-\chi_j)$ and $N^{-1}\sum_ie^{i\theta_i}=N^{-1}\sum_ie^{i(\theta'_i+\chi_i)}$.
Dropping the primes, we obtain
\begin{align}
&
\mathbb E_{\mathrm{corr}}
\left[
\left\langle
\psi(M_{\Lambda_L})
\right\rangle_{\mathrm{corr},h}
\right]
\nonumber\\
={}&
\int d\mu(B)\,d\rho(r)\,d\nu(\phi)\,
\frac{
\exp\left[
\beta\sum_{\langle ij\rangle}
\cos(B_{ij}+\chi_i-\chi_j)
\right]
}{Z_0(B;\beta)}
\nonumber\\
&\times
\frac{
\displaystyle
\int d\nu(\theta)\,
\psi
\left(
\frac{1}{N}\sum_i e^{i(\theta_i+\chi_i)}
\right)
e^{-\beta H_{\mathrm{corr}}(h)}
}{
Z_h(B,r,\phi)
}.
\label{eq:gauge_transformed_distribution}
\end{align}

Since this equality holds for arbitrary
site angles $\chi_i$, we may average both sides
over independent uniform angles with $d\nu(\chi)$.
We then introduce two spin configurations
\begin{equation}
\theta_i^{(1)}=\theta_i,
\qquad
\theta_i^{(2)}=-\chi_i,
\end{equation}
for which
\begin{align}
\cos(B_{ij}+\chi_i-\chi_j)
&=
\cos(\theta_i^{(2)}-\theta_j^{(2)}-B_{ij}),
\nonumber\\
\frac{1}{N}\sum_i e^{i(\theta_i+\chi_i)}
&=
Q_{\Lambda_L}.
\end{align}
Therefore,
\begin{align}
&
\mathbb E_{\mathrm{corr}}
\left[
\left\langle
\psi(M_{\Lambda_L})
\right\rangle_{\mathrm{corr},h}
\right]
\nonumber\\
={}&
\int d\mu(B)\,d\rho(r)\,d\nu(\phi)
\int d\nu(\theta^{(1)})\,d\nu(\theta^{(2)})
\nonumber\\
&\times
\frac{
\exp\left[
\beta\sum_{\langle ij\rangle}
\cos(\theta_i^{(2)}-\theta_j^{(2)}-B_{ij})
\right]
}{Z_0(B;\beta)}
\nonumber\\
&\times
\frac{
e^{-\beta H_{\mathrm{corr}}(h)}
}{
Z_h(B,r,\phi)
}\,
\psi(Q_{\Lambda_L}),
\label{eq:two_replica_integral}
\end{align}
where the Hamiltonian in the second Gibbs weight
is evaluated at $(\theta^{(1)},B,r,\phi)$.
The first Gibbs weight in
Eq.~\eqref{eq:two_replica_integral}
corresponds to the zero-field XY gauge-glass model,
while the second corresponds to the Gaussian
random-field XY gauge-glass model.
Moreover, the bond variables $B_{ij}$ are now
integrated with respect to the independent
uniform distribution $d\mu(B)$.
Thus, $\theta^{(1)}$ and $\theta^{(2)}$
represent independent equilibrium spin configurations
of the model~\eqref{eq:random_field_polar}
at field strengths $h$ and $0$, respectively,
sampled with the same realization of the bond disorder, $A_{ij}=B_{ij}$.
We therefore obtain
\begin{align}
&
\mathbb E_{\mathrm{corr}}
\left[
\left\langle
\psi(M_{\Lambda_L})
\right\rangle_{\mathrm{corr},h}
\right]
=
\mathbb E_{\mathrm{RF}}
\left[
\left\langle
\psi(Q_{\Lambda_L})
\right\rangle_{\mathrm{RF};h,0}
\right]
\label{eq:overlap_magnetization_identity}
\end{align}
for every finite system size.
Combining this result with the first equality
establishes Eqs.~\eqref{eq:main_identity} and \eqref{eq:identity_psi}.

%%%%%%%%%%%%%%%%%%%%%%%%%%%%%%%%%%%%%%%%%%%%%%%%%%%%%%%%%%%%%%%
%%%%%%%%%%%%%%%%%%%%%%%%%%%%%%%%%%%%%%%%%%%%%%%%%%%%%%%%%%%%%%%
\subsection{Reduction to the absence of ferromagnetic order}\label{sec:reduction}

Equation~\eqref{eq:identity_psi} with $\psi(q)=|q|^2$ gives
\begin{equation}
\mathbb E_{\mathrm F}\left[\left\langle|Q_{\Lambda_L}|^2\right\rangle_{\mathrm F;h,0}\right]
=
\mathbb E_{\mathrm{corr}}\left[\left\langle|M_{\Lambda_L}|^2\right\rangle_{\mathrm{corr},h}\right]
\label{eq:QM_second_moment}
\end{equation}
for every $L$.
This identity establishes an exact
connection between the response of the XY gauge glass
to an applied magnetic field and ferromagnetic order
in the correlated XY gauge-glass model subjected
to a random magnetic field.
At $h=0$, it further implies that the second moment of the magnetization of the correlated model equals that of the spin-glass overlap of the XY gauge glass, as in the Ising case~\cite{Nishimori2024}.

Our goal is to show that, for any finite inverse
temperature $\beta>0$ and any $h>0$,
the overlap vanishes in the sense of Eq.~\eqref{eq:second_moment}, i.e.,
$P_{\mathrm F}(q;0,h)=\delta^{(2)}(q)$ for $d\leq4$.
By Eq.~\eqref{eq:QM_second_moment}, this is equivalent to
\begin{equation}
\lim_{L\to\infty}
\mathbb E_{\mathrm{corr}}\left[\left\langle|M_{\Lambda_L}|^2\right\rangle_{\mathrm{corr},h}\right]=0
\qquad (d\leq4),
\label{eq:required_random_field_bound}
\end{equation}
that is, $P_{\mathrm{corr}}(m;h)=\delta^{(2)}(m)$.
Thus, the original problem reduces to proving
the absence of ferromagnetic long-range order
in the correlated XY gauge-glass model
under a random magnetic field.

The existence or absence of ferromagnetic order
in continuous-spin systems subjected to random fields
has been studied extensively since the pioneering
work of Imry and Ma~\cite{Imry1975}.
Using an energy-balance argument, they predicted
that ferromagnetic long-range order cannot survive
in the random-field XY model in dimensions $d\leq4$:
reorienting the spins in a region of linear size $\ell$ costs an energy of order $\ell^{d-2}$ for continuous spins, since the reorientation can be spread smoothly over the region as a spin wave, whereas the random-field energy that can be gained is of order $h\ell^{d/2}$. The latter dominates for $d<4$, and $d=4$ is the marginal case.
Aizenman and Wehr~\cite{AizenmanWehr1989,AizenmanWehr1990}
subsequently established this
prediction rigorously for a broad class of classical
spin systems with continuous symmetry and
translation-invariant interactions; the result has also been extended to quantum systems~\cite{AGL2012}.

Their theorem does not directly apply to the correlated model, however. The couplings $B_{ij}$ are random and correlated, and what is needed is a bound valid for each realization of the bonds and uniform in the realization.
The quantitative estimates of Dario, Harel, and Peled~\cite{Dario2024}
(DHP hereafter) on the magnetization of classical
continuous-spin systems subjected to Gaussian random
fields in dimensions $d\leq4$ provide the starting point for such a bound.
The DHP estimates are averages over the Gaussian random field only, for a given deterministic interaction, and the constants depend only on uniform bounds on the interaction. The fixed-bond estimates required here are established in Appendix~\ref{app:proof} by adapting their proof to bond-dependent interactions.
Since the random field is independent of the bonds under $\mathbb E_{\mathrm{corr}}$, we can write
\begin{equation}
\mathbb E_{\mathrm{corr}}[X]
=\int d\mu(B)\,
\frac{e^{\beta\sum_{\langle ij\rangle}\cos B_{ij}}}{Z_0(B;\beta)}\,
\E_\eta[X],
\label{eq:Ecorr_decomposition}
\end{equation}
apply the estimates for a fixed bond configuration $B$ in the inner average $\E_\eta$, and average over the correlated distribution afterwards. The correlations among the bonds and the absence of translation invariance for a given $B$ then play no role.

The assumptions of DHP, namely an $O(n)$-invariant interaction that does not depend on the bond and translation invariance in the step that treats the periodic system, are not literally satisfied in the present model. In Appendix~\ref{app:proof}, we therefore give a proof adapted to the present model, which is self-contained for $d\leq3$; for $d=4$, the key estimates of the multiscale argument and the treatment of general box sizes are derived in condensed form, and further details of the covering construction can be found in DHP. This yields the following result.

%%%%%%%%%%%%%%%%%%%%%%%%%%%%%%%%%%%%%%%%%%%%%%%%%%%%%%%%%%%%%%%
\begin{theorem}[Absence of ferromagnetic order]\label{thm:main}
Let $d\leq4$ and $h>0$. There exists a sequence $\varepsilon_L$, depending only on $d$, $h$, and $L$ and satisfying $\varepsilon_L\to0$ as $L\to\infty$, with the following properties, valid for every inverse temperature $\beta>0$ and every bond configuration $B=\{B_{ij}\}$:
(i) for the model~\eqref{eq:correlated_model} with fixed bonds $B$ and periodic boundary conditions,
\begin{equation}
\E_\eta\left[\left\langle|M_{\Lambda_L}|^2\right\rangle_{\mathrm{corr},h}\right]\leq\varepsilon_L ;
\label{eq:theorem_second_moment}
\end{equation}
(ii) for the same model on a box $\Lambda_L$ of linear size $L$, instead of the torus, with boundary condition $\tau$ (the angles of the spins outside $\Lambda_L$ adjacent to it),
\begin{equation}
\E_\eta\left[\sup_\tau\left|\left\langle M_{\Lambda_L}\right\rangle^{\tau}_{\mathrm{corr},h}\right|\right]\leq\varepsilon_L ,
\label{eq:theorem_sup}
\end{equation}
where the supremum is taken over all boundary conditions, including those depending on the random field.
One can take $\varepsilon_L=C L^{-(4-d)/[4(8-d)]}$ for $d\leq 3$ and $\varepsilon_L=\min\{1,C/\sqrt{\ln\ln L}\}$ for $d=4$ and $L\geq16$ ($\varepsilon_L=1$ for $d=4$ and $L<16$), with $C$ depending only on $d$ and $h$. (For $d\leq3$, the bound~(ii) in fact holds with the larger exponent $(4-d)/[2(8-d)]$.)
In particular, the correlated XY gauge-glass model
defined by Eq.~\eqref{eq:correlated_model} satisfies Eq.~\eqref{eq:required_random_field_bound}, i.e.,
\begin{equation}
P_{\mathrm{corr}}(m;h)
=\delta^{(2)}(m)
\qquad (d\leq4).
\label{eq:conjecture}
\end{equation}
\end{theorem}
%%%%%%%%%%%%%%%%%%%%%%%%%%%%%%%%%%%%%%%%%%%%%%%%%%%%%%%%%%%%%%%

The last statement follows from Eqs.~\eqref{eq:Ecorr_decomposition} and \eqref{eq:theorem_second_moment}, since the bound~\eqref{eq:theorem_second_moment} holds for every $B$ with the same $\varepsilon_L$. Theorem~\ref{thm:main} controls the second moment $\langle|M_{\Lambda_L}|^2\rangle$, which is needed for the distribution function, rather than the thermal average $|\langle M_{\Lambda_L}\rangle|$ bounded by DHP. The step from the latter to the former, i.e., from (ii) to (i), uses the consistency of Gibbs distributions (the DLR property) and is given in Appendix~\ref{app:secondmoment}.

Theorem~\ref{thm:main} is a consequence of the Imry--Ma mechanism, which does not depend on the details of the bonds. It applies equally to the pure XY ferromagnet ($B_{ij}=0$) and to any fixed realization of $B$. The reason is that the spin-wave estimate compares each configuration with the same configuration twisted by a slowly varying angle. Since the bond energy $-\cos(\theta_i-\theta_j-B_{ij})$ depends only on the angle difference, the twist changes it by an amount controlled by the change of the twist angle across the bond, of order $1/\ell$, whatever the values of $\theta_i-\theta_j$ and $B_{ij}$. After the first-order terms are cancelled by combining twists in opposite directions, the cost is bounded by a constant times $\ell^{d-2}$ for every configuration, however irregular (Appendix~\ref{app:spinwave}). The frustration of the bonds and the irregularity of the spin configurations therefore do not affect this bound, and the Imry--Ma energy balance is the same as for the pure XY ferromagnet. The conclusion of Corollary~\ref{cor:main} below therefore holds irrespective of whether a spin-glass phase exists and of its nature. The gauge identity~\eqref{eq:main_identity} translates a general property of continuous-spin systems in $d\leq 4$ into a statement about the response of the spin-glass phase to a magnetic field.

We also note that Theorem~\ref{thm:main} concerns the absence of long-range order. Whether the random-field XY model in three dimensions has quasi-long-range order, with a power-law decay of correlations, is a separate question that has been discussed in the literature~\cite{GingrasHuse1996,Feldman2000,TissierTarjus2006}.

Combining Theorem~\ref{thm:main} with the exact identity~\eqref{eq:QM_second_moment}, we immediately
obtain the following corollary.

%%%%%%%%%%%%%%%%%%%%%%%%%%%%%%%%%%%%%%%%%%%%%%%%%%%%%%%%%%%%%%%
\begin{corollary}\label{cor:main}
Let $d\leq4$. For any finite inverse temperature $\beta>0$
and any $h>0$, the XY gauge-glass model
defined by Eq.~\eqref{eq:biased_field} satisfies
\begin{equation}
\lim_{L\to\infty}
\mathbb E_{\mathrm F}\left[\left\langle|Q_{\Lambda_L}|^2\right\rangle_{\mathrm F;h,0}\right]=0,
\end{equation}
that is,
\begin{equation}
P_{\mathrm F}(q;0,h)
=\delta^{(2)}(q)
\qquad (d\leq4).
\label{main-result}
\end{equation}
\end{corollary}

Corollary~\ref{cor:main} concerns the disorder average. Since $\langle|Q_{\Lambda_L}|^2\rangle_{\mathrm F;h,0}\geq0$ for each sample, it implies that the probability, with respect to the disorder, that $\langle|Q_{\Lambda_L}|^2\rangle_{\mathrm F;h,0}$ exceeds any fixed $\epsilon>0$ tends to zero as $L\to\infty$. The overlap therefore vanishes for typical samples. We do not claim that it vanishes almost surely as $L\to\infty$.
%%%%%%%%%%%%%%%%%%%%%%%%%%%%%%%%%%%%%%%%%%%%%%%%%%%%%%%%%%%%%%%

%%%%%%%%%%%%%%%%%%%%%%%%%%%%%%%%%%%%%%%%%%%%%%%%%%%%%%%%%%%%%%%
\subsection{Implications for magnetic-field chaos}\label{sec:fieldchaos}

We now discuss the physical implications of
Eq.~\eqref{main-result} for magnetic-field chaos.

According to the RSB
picture, the spin-glass phase can remain stable
under a sufficiently weak magnetic field.
Nevertheless, the field induces a drastic
rearrangement of equilibrium spin configurations,
causing configurations sampled at zero and
nonzero fields to lose their mutual overlap in the thermodynamic limit~\cite{Kondor1989,Ritort1994,Billoire2003,AguilarJanita2024}.
This phenomenon, known as magnetic-field chaos,
is characterized within the spin-glass phase by
\begin{equation}
P_{\mathrm F}(q;0,h)=\delta^{(2)}(q)
\qquad (h>0).
\label{eq:rsb_field_chaos}
\end{equation}
In contrast, the droplet picture predicts that
an arbitrarily weak magnetic field destroys
the spin-glass phase in the thermodynamic limit~\cite{Fisher1986,Fisher1988}.
In this scenario, the overlap between equilibrium configurations
at zero and nonzero fields is also expected
to vanish, leading to
\begin{equation}
P_{\mathrm F}(q;0,h)=\delta^{(2)}(q)
\qquad (h>0).
\end{equation}
Thus, the vanishing of the cross-overlap
in Eq.~\eqref{main-result} is consistent
with both the RSB and droplet pictures
and does not, by itself, distinguish
between these two scenarios.

There is, however, a third logically possible
scenario in which the spin-glass phase persists
under a magnetic field without exhibiting
magnetic-field chaos.
In this case, equilibrium configurations
at zero and nonzero fields would retain
a nonvanishing overlap in the thermodynamic limit:
\begin{equation}
\limsup_{L\to\infty}
\mathbb E_{\mathrm F}\left[\left\langle|Q_{\Lambda_L}|^2\right\rangle_{\mathrm F;h,0}\right]>0
\qquad (h>0),
\label{eq:no_field_chaos}
\end{equation}
that is, $P^{(L)}_{\mathrm F}(q;0,h)$ does not converge to $\delta^{(2)}(q)$: at least one subsequence has a limit different from $\delta^{(2)}(q)$.
By the rotational symmetry of $P_{\mathrm F}^{(L)}$ noted above, Eq.~\eqref{eq:no_field_chaos} means that the distribution of $|q|$ does not concentrate at the origin.
Our result in Eq.~\eqref{main-result}
rigorously rules out this possibility
in dimensions $d\leq4$.

More precisely, our result does not rely on either the RSB or the droplet picture. Even if neither picture correctly describes the XY gauge-glass model in three or four dimensions, it follows that, whenever a spin-glass phase exists at zero field, any nonzero magnetic field must either induce magnetic-field chaos in the surviving phase or destroy spin-glass order altogether.
In particular, if the RSB picture is correct in the sense that the spin-glass phase survives in a sufficiently weak field
of the present type, Eq. \eqref{main-result} shows that the surviving phase exhibits magnetic-field chaos in the sense of Eq. \eqref{eq:rsb_field_chaos}.
In mean-field theory, the stability of the spin-glass phase in a field and magnetic-field chaos both follow from the
RSB solution; in finite dimensions they are logically distinct properties, and our result shows that, for the present
model in $d\le 4$, the former implies the latter.

We emphasize that our theorem concerns the overlap between equilibrium configurations at zero field and at nonzero field. It neither establishes nor excludes a spin-glass phase at nonzero field; characterizing such a phase would require other quantities, such as the overlap between two replicas in the same field, connected correlation functions, or the spin-glass susceptibility. The statement about magnetic-field chaos above is a consequence of the theorem under the condition that a spin-glass phase exists. We also note that the vanishing of the overlap is not, by itself, specific to the spin-glass phase; it holds trivially in the paramagnetic phase. Finally, $Q_{\Lambda_L}$ is the standard complex overlap of the gauge glass, which is invariant under the gauge transformation. Other combinations of the spin components of the two replicas, such as $N^{-1}\sum_ie^{i(\theta_i^{(1)}+\theta_i^{(2)})}$, are not covered by our result.

We note the order of limits. Our results hold for each fixed $h>0$ in the limit $L\to\infty$, and the limit $h\to0$ is to be taken afterwards. At $h=0$ itself, $P_{\mathrm F}(q;0,0)$ is the zero-field overlap distribution, which is nontrivial in a spin-glass phase. In a spin-glass phase, the limit $h\to0^+$ of $P_{\mathrm F}(q;0,h)$ therefore differs from $P_{\mathrm F}(q;0,0)$.

%%%%%%%%%%%%%%%%%%%%%%%%%%%%%%%%%%%%%%%%%%%%%%%%%%%%%%%%%%%%%%%

%%%%%%%%%%%%%%%%%%%%%%%%%%%%%%%%%%%%%%%%%%%%%%%%%%%%%%%%%%%%%%%
\subsection{Generalization to different temperatures}\label{sec:twoT}
Finally, we extend our result to the case of two equilibrium configurations at different temperatures.
Let us denote the parameter in the bond distribution~\eqref{eq:Pcorr} by $\beta_p$, so that $P_{\mathrm{corr}}(B)$ and $Z_0(B;\beta_p)$ involve $\beta_p$, while the Gibbs weight of the correlated model~\eqref{eq:correlated_model} is $e^{-\beta H_{\mathrm{corr}}(h)}$. 
The derivation of Eq.~\eqref{eq:overlap_magnetization_identity} goes through without change, and yields
\begin{align}
&\mathbb E_{\mathrm{corr}}^{(\beta_p)}
\left[\left\langle\psi(M_{\Lambda_L})\right\rangle_{\mathrm{corr};\beta,h}\right]
\nonumber\\
&=\mathbb E_{\mathrm{RF}}
\left[\left\langle\psi(Q_{\Lambda_L})\right\rangle_{\mathrm{RF};(\beta,h),(\beta_p,0)}\right]
\nonumber\\
&=\mathbb E_{\mathrm{F}}
\left[\left\langle\psi(Q_{\Lambda_L})\right\rangle_{\mathrm{F};(\beta,h),(\beta_p,0)}\right],
\label{eq:identity_twoT}
\end{align}
where $\mathbb E_{\mathrm{corr}}^{(\beta_p)}$ is defined by Eq.~\eqref{eq:disorder_averages} with $\beta$ replaced by $\beta_p$, and $\langle\cdot\rangle_{\mathsf M;(\beta,h),(\beta_p,0)}$ denotes the two-replica average with the first replica at inverse temperature $\beta$ and field $h$ and the second at inverse temperature $\beta_p$ and zero field. The last equality follows from the change of variables~\eqref{eq:gauge_transformation}, which does not involve the temperature. Since Theorem~\ref{thm:main} holds for every $\beta$ and every $B$, we obtain the following.

\begin{corollary}\label{cor:twoT}
Let $d\leq4$. For any $\beta_1,\beta_2>0$ and any $h>0$,
\begin{equation}
\lim_{L\to\infty}
\mathbb E_{\mathrm F}\left[\left\langle|Q_{\Lambda_L}|^2\right\rangle_{\mathrm F;(\beta_2,h),(\beta_1,0)}\right]=0
\qquad (d\leq4).
\end{equation}
\end{corollary}

Thus, the overlap between equilibrium configurations at zero field and at nonzero field vanishes even when their temperatures differ; this includes simultaneous changes of temperature and field. The case in which both replicas are in nonzero fields, $h_1,h_2>0$, is not covered. The gauge transformation then introduces the random field into the bond distribution, so that the field and the bonds become correlated, and Theorem~\ref{thm:main} cannot be applied in the form~\eqref{eq:Ecorr_decomposition}. Indeed, the overlap does not vanish in general in this case, because both replicas are polarized along the field direction; see Ref.~\cite{Nishimori2025} for the Ising case.

The case of zero field in both replicas, $h=0$ with $\beta\neq\beta_p$, is not covered either. In this case, Eq.~\eqref{eq:identity_twoT} equates the overlap between zero-field configurations at different temperatures, which characterizes temperature chaos, with the magnetization of the correlated model at $\beta\neq\beta_p$ without a random field~\cite{Nishimori2024,Nishimori2025a}. The Imry--Ma mechanism is then not available, and whether this magnetization vanishes is precisely the question of temperature chaos.

%%%%%%%%%%%%%%%%%%%%%%%%%%%%%%%%%%%%%%%%%%%%%%%%%%%%%%%%%%%%%%%

\section{Discussion}\label{sec:discussion}
We have proved that, in the XY gauge-glass model defined by Eq.~\eqref{eq:biased_field}, the disorder average of the squared spin overlap between equilibrium configurations at zero and nonzero magnetic fields vanishes in the thermodynamic limit at any finite temperature in dimensions $d\leq 4$, and that the same holds when the two configurations are at different temperatures. This result rules out the possibility of a spin-glass phase that persists under a magnetic field without exhibiting magnetic-field chaos. Our result thus provides one of the few rigorous constraints on the nature of spin-glass phases in finite-dimensional systems.

Our result follows from the Imry--Ma mechanism, which suppresses magnetization in the correlated model~\eqref{eq:correlated_model} for $d\leq 4$. 
The exact identity~\eqref{eq:QM_second_moment}, derived from gauge symmetry, translates this property into a statement about the spin overlap between equilibrium states of the XY gauge glass model~\eqref{eq:biased_field} at zero and nonzero magnetic fields.
The result therefore holds independently of the existence and the properties of the spin-glass phase; if a spin-glass phase exists, the result constrains the response of this phase to a magnetic field.

The exact identity in Eq.~\eqref{eq:QM_second_moment} underlying our result can also be derived for other spin-glass models with gauge symmetry, including the Ising spin-glass model~\cite{Nishimori2025}. However, the corresponding rigorous results on the absence of long-range order in random-field systems with discrete spins are restricted to dimensions $d\leq 2$~\cite{AizenmanWehr1989,AizenmanWehr1990,Chatterjee2018,AizenmanPeled2019,AizenmanHarelPeled2020,DingXia2021}; in $d\geq3$, the random-field Ising model retains long-range order at low temperatures and weak disorder~\cite{Imbrie1985,BricmontKupiainen1988,DingZhuang2024}.
Since the two-dimensional Ising spin-glass model is already believed to exhibit no spin-glass order at any finite temperature~\cite{Hartmann2001,Jorg2006,Ohzeki2009}, applying our approach to this model does not yield a nontrivial constraint on the stability of a finite-temperature spin-glass phase.

In this work, we have considered Gaussian random fields
for technical reasons.
Another natural choice is a random field of fixed magnitude
whose direction is uniformly distributed on the circle.
The corresponding Hamiltonian of the correlated
XY gauge-glass model is given by
\begin{equation}
H_{\mathrm{corr}}(h)
=
-\sum_{\langle ij\rangle}
\cos(\theta_i-\theta_j-B_{ij})
-h\sum_i \cos(\theta_i-\phi_i).
\end{equation}
This random-field distribution also possesses gauge symmetry,
and the same gauge-transformation argument yields
an exact identity analogous to Eq.~\eqref{eq:main_identity}.
In this case, the corresponding conventional
XY gauge-glass model is subjected to a uniform magnetic field,
with Hamiltonian
\begin{equation}
H_{\mathrm{F}}(h)
=
-\sum_{\langle ij\rangle}
\cos(\theta_i-\theta_j-A_{ij})
-h\sum_i \cos\theta_i.
\end{equation}
Although we expect an analogue of Corollary \ref{cor:main} to hold in dimensions $d\leq 4$, establishing it rigorously would require additional technical arguments, and we therefore do not claim such a result here.
The method of DHP, which we adapt to obtain such bounds, is formulated for Gaussian fields~\cite{Dario2024}.
In its proof, the Gaussian distribution is used in an essential way: the field averaged over a box is a Gaussian variable independent of the remaining components of the field (Appendices~\ref{app:setting} and \ref{app:fixedbc}). For fields of fixed magnitude, the spin-wave bound on the free energy (Appendix~\ref{app:spinwave}) still holds, since it requires only the rotational invariance and the independence of the fields at different sites, but the subsequent steps would have to be modified.

One might hope to overcome this limitation by applying the Aizenman--Wehr theorem~\cite{AizenmanWehr1989,AizenmanWehr1990}, which covers a broad class of field distributions. However, its result for systems with continuous symmetry (Theorem~4.4 of Ref.~\cite{AizenmanWehr1990}) requires the field distribution to be either absolutely continuous with respect to Lebesgue measure or, at positive temperatures, spherically symmetric with no isolated point masses in its radial distribution.
A field of fixed magnitude, whose radial distribution is concentrated at a single point, satisfies neither condition. Moreover, the theorem is formulated for translation-invariant interactions and does not provide bounds for each fixed realization of the bonds, which we need.

The bounds in Theorem~\ref{thm:main} are uniform in the inverse temperature $\beta$, and DHP also give zero-temperature versions of their estimates. This suggests that a zero-temperature version of our result holds, for instance for the overlap between the ground state of Eq.~\eqref{eq:biased_field} at $h>0$ and zero-field equilibrium configurations at a finite temperature, obtained from Eq.~\eqref{eq:identity_twoT} by letting $\beta\to\infty$ at fixed $\beta_p$. A careful treatment of this limit, and of the case in which both replicas are at zero temperature, which requires defining the limit $\beta_p\to\infty$ of the correlated distribution, is left for future work.

We emphasize that the identity~\eqref{eq:QM_second_moment} is an equality in any dimension, so that the vanishing of the overlap is equivalent to the absence of ferromagnetic order in the correlated model under the random field. As noted in Sec.~\ref{sec:reduction}, the correlated model is ferromagnetic at zero field precisely when the XY gauge glass is in the spin-glass phase.

A spin-glass phase is widely believed to exist at zero field in sufficiently high dimensions, and both the RSB and droplet pictures predict that the overlap between equilibrium configurations at zero and nonzero fields vanishes there (Sec.~\ref{sec:fieldchaos}). If this is the case, the identity implies that the ferromagnetic order of the correlated model at zero field is destroyed by an arbitrarily weak random field also in $d\geq5$. This is the counterpart of the conclusion drawn for the Ising spin glass with correlated disorder in Ref.~\cite{Nishimori2025}, where the ferromagnetic phase was shown to be unstable under random fields in any dimension if magnetic-field chaos exists. It is in sharp contrast to the pure XY ferromagnet, which is expected to retain its order under a weak random field for $d\geq5$~\cite{Imry1975}. The Imry--Ma mechanism is not available there, and the destruction of order would instead reflect the sensitivity of the spin-glass state to a magnetic field, inherited by the correlated bond distribution through gauge symmetry.

The destruction of order in $d\geq5$ can be understood heuristically by an energy balance of the Imry--Ma type in which the spin-wave cost is replaced by the free-energy cost of spin-glass droplets. Since the frustration, i.e., the sum of $B_{ij}$ around each plaquette modulo $2\pi$, is invariant under $B_{ij}\to B_{ij}+\chi_i-\chi_j$, the argument leading to Eq.~\eqref{eq:Pcorr_normalization} shows that its statistics in the correlated model are the same as in the conventional XY gauge glass. The ordered state of the correlated model therefore competes not only with smoothly rotated states, as in the pure XY ferromagnet, but also with other spin-glass states. Suppose that the zero-field XY gauge glass is described by droplets whose free-energy cost scales as $\ell^{\theta}$ with linear size $\ell$, and that the reorientation of a droplet involves a finite fraction of its spins. A random field independent of the zero-field states then produces an energy difference of typical size $h\ell^{d/2}$. The field therefore wins on large scales in any dimension if $\theta<d/2$, which would follow from the bound $\theta\leq(d-1)/2$ derived by Fisher and Huse for Ising spin glasses~\cite{Fisher1988} if the same bound held for the present XY model; we do not establish this. The conclusion is consistent with the droplet prediction that a field destroys the spin-glass phase; in the RSB picture, the same conclusion follows from the magnetic-field chaos predicted by mean-field theory.

From this point of view, the present proof for $d\leq4$ uses only the spin-wave bound $\ell^{d-2}$, which holds for any configuration, whereas $d\geq5$ would require the actual stiffness of the spin-glass state, which is precisely the issue between the droplet and RSB pictures. A proof in $d\geq5$ is therefore beyond the present method.

Conversely, ferromagnetic order of the correlated model under a weak random field in some $d\geq5$ would imply a nonvanishing overlap between equilibrium configurations of the conventional XY gauge glass at zero and nonzero fields, i.e., Eq.~\eqref{eq:no_field_chaos}. If, in addition, a spin-glass phase at the nonzero field were established independently, this would be the third scenario of Sec.~\ref{sec:fieldchaos}, a spin-glass phase that persists under a magnetic field without field chaos. Numerical studies of the correlated model in $d\geq5$ would therefore provide a test of magnetic-field chaos in terms of the magnetization of a single system rather than the overlap between two replicas at different fields. 
For such studies, bond configurations distributed according to Eq.~\eqref{eq:Pcorr} can be generated directly from equilibrium configurations of the conventional XY gauge glass, without computing $Z_0(B;\beta)$, using the method proposed by Oshima, Arai, and Hukushima~\cite{Oshima2026} for a two-dimensional Ising spin-glass model with correlated disorder ($B_{ij}=A_{ij}-\theta_i+\theta_j$, with uniformly distributed $A_{ij}$ and $\theta$ an equilibrium configuration of the conventional model with bonds $A_{ij}$ at zero field and inverse temperature $\beta$).
The preparation of such equilibrium configurations at low temperatures, however, remains as difficult as in direct simulations of the conventional model.

%%%%%%%%%%%%%%%%%%%%%%%%%%%%%%%%%%%%%%%%%%%%%%%%%%%%%%%%%%%%%%%
\section*{Methods: Use of generative-AI tools}\label{sec:methods}

Generative-AI tools were used extensively as research assistants in carrying out the analytical calculations reported in this work. The authors determined the physical questions to be addressed, the detailed strategy of the analysis, and the calculations to be performed. Claude Opus 5.5 and Claude Fable 5.1 (Anthropic) were used primarily to carry out analytical derivations, including the adaptation of the estimates of Ref.~\cite{Dario2024}
to the present model in Appendix~\ref{app:proof}, under detailed instructions from the authors.

For the analytical calculations, the authors examined and independently verified each step of the AI-generated derivations. GPT-6.1 Sol (OpenAI) and DeepReview (Google) were additionally used to independently examine the derivations and identify possible errors or inconsistencies. Issues found either by the authors or through this cross-checking process were fed back to the Claude models for correction, after which the revised calculations were again examined by the authors. This iterative procedure was repeated throughout the analytical development.

All decisions concerning the physical approach, the acceptance or rejection of AI-generated results, and their interpretation were made by the authors, who take full responsibility for the scientific content of this work.

%%%%%%%%%%%%%%%%%%%%%%%%%%%%%%%%%%%%%%%%%%%%%%%%%%%%%%%%%%%%%%%%%%
\begin{acknowledgments}
M. Okuyama was supported by JST BOOST (Grant No. JPMJBY24B6) and JSPS KAKENHI (Grant No. 24K16973). M. Ohzeki was supported by the Cross-ministerial Strategic Innovation Promotion Program (SIP) of the Cabinet Office, Japan (Grant No. 23836436).

Generative-AI tools, Claude Opus 5.5 and Claude Fable 5.1 (Anthropic), with additional use of GPT-6.1 Sol (OpenAI), were used extensively in preparing the manuscript through iterative interaction with the authors (their use in the analysis is described in the Methods section). The authors provided detailed and repeated instructions for revision, correction, and restructuring, and reviewed and edited the resulting text. The authors take full responsibility for the final wording and presentation of the manuscript.

\end{acknowledgments}

%%%%%%%%%%%%%%%%%%%%%%%%%%%%%%%%%%%%%%%%%%%%%%%%%%%%%%%%%%%%%%%
\section*{Data availability statement}
No new data were created or analysed in this study.

%%%%%%%%%%%%%%%%%%%%%%%%%%%%%%%%%%%%%%%%%%%%%%%%%%%%%%%%%%%%%%%%%%
\appendix
%%%%%%%%%%%%%%%%%%%%%%%%%%%%%%%%%%%%%%%%%%%%%%%%%%%%%%%%%%%%%%%%%%
\section{Proof of Theorem~\ref{thm:main}}\label{app:proof}

In this Appendix, we prove Theorem~\ref{thm:main}. The proof follows Sec.~8 of DHP~\cite{Dario2024},
adapted to the present model, which has two spin components, bond-dependent couplings, and no translation invariance for a fixed bond configuration. We have tried to keep the presentation self-contained. The exception is the multiscale covering in four dimensions (Sec.~\ref{app:d4}): its key estimates and the treatment of general box sizes are described in condensed form, while further details of the covering construction are taken from DHP, and we explain why they apply without change. The points that would require more detail in a fully rigorous treatment, and the correspondence with the statements of DHP, are listed in Sec.~\ref{app:remarks}.

\subsection{Setting and basic properties}\label{app:setting}

Throughout this Appendix, the bond variables $B=\{B_{ij}\}$ are fixed but arbitrary, and $\beta>0$ and $h>0$ are fixed. We write $\bsig_i=(\cos\theta_i,\sin\theta_i)$ and $\bet_i=(\eta_i^x,\eta_i^y)$, and denote their components by $\sigma_{i,a}$ and $\eta_{i,a}$ with $a\in\{x,y\}$.

For a finite set of sites $\Lambda$ (a box or a torus), a boundary condition $\tau=\{\tau_j\}$ specifies the angles of the spins on the outer boundary $\partial\Lambda$, i.e., the sites outside $\Lambda$ adjacent to $\Lambda$.
We add an auxiliary uniform field $\bu\in\mathbb R^2$, which is used only as a device in Sec.~\ref{app:uniform}, and consider
\begin{align}
H^{\tau}_{\Lambda}
={}&-\sum_{\langle ij\rangle;\,i,j\in\Lambda}\cos(\theta_i-\theta_j-B_{ij})
\nonumber\\
&-\sum_{\langle ij\rangle;\,i\in\Lambda,\,j\in\partial\Lambda}\cos(\theta_i-\tau_j-B_{ij})
\nonumber\\
&-\sum_{i\in\Lambda}(h\bet_i+\bu)\cdot\bsig_i .
\label{eq:A_hamiltonian}
\end{align}
We also use periodic boundary conditions, denoted by $\tau=\mathrm{per}$, with arbitrary values of $B_{ij}$ on the bonds that wrap around the torus.
The thermal average, the partition function, and the free energy per site are
\begin{align}
\langle O\rangle^{\tau}_{\Lambda}
&=\frac{1}{Z^\tau_\Lambda}\int\prod_{i\in\Lambda}\frac{d\theta_i}{2\pi}\,O\,e^{-\beta H^{\tau}_{\Lambda}},
\nonumber\\
Z^\tau_\Lambda(\bet,\bu)
&=\int\prod_{i\in\Lambda}\frac{d\theta_i}{2\pi}\,e^{-\beta H^{\tau}_{\Lambda}},
\nonumber\\
f^\tau_\Lambda(\bet,\bu)
&=-\frac{1}{\beta|\Lambda|}\ln Z^\tau_\Lambda(\bet,\bu),
\label{eq:A_free_energy}
\end{align}
where $|\Lambda|$ is the number of sites in $\Lambda$.
When $\Lambda_0$ denotes the whole system and $\Lambda\subseteq\Lambda_0$ is a box, the magnetization of $\Lambda$ in direction $a$ is
\begin{equation}
m^{\tau}_{\Lambda,a}
=\frac{1}{|\Lambda|}\sum_{i\in\Lambda}\langle\sigma_{i,a}\rangle^{\tau}_{\Lambda_0}.
\label{eq:A_magnetization}
\end{equation}
When $\Lambda$ itself is the whole system, $m^{\tau}_{\Lambda,a}$ is its magnetization with the boundary condition $\tau$.
The symbol $\E_\eta$ denotes the average~\eqref{eq:Eeta} over the Gaussian field, with $B$, $\beta$, $h$, and $\bu$ fixed.
For a box $\Lambda$ and a component $a$, we decompose $\eta_{i,a}=\heta_{\Lambda,a}+\eta^{\perp}_{i,a}$ for $i\in\Lambda$, where
\begin{equation}
\heta_{\Lambda,a}=\frac{1}{|\Lambda|}\sum_{i\in\Lambda}\eta_{i,a}.
\end{equation}
The box-averaged field $\heta_{\Lambda,a}$ is a Gaussian variable with variance $1/|\Lambda|$, independent of $\{\eta^{\perp}_{i,a}\}$ and of all the other field variables. This independence is a property of the Gaussian distribution.

We use the following elementary properties.

(P1) \textit{Derivatives and concavity.}
Since the fields enter the Hamiltonian linearly,
\begin{equation}
\frac{\partial f^\tau_{\Lambda_0}}{\partial\heta_{\Lambda,a}}=-\frac{h|\Lambda|}{|\Lambda_0|}\,m^\tau_{\Lambda,a},
\qquad
\frac{\partial f^\tau_{\Lambda_0}}{\partial u_a}=-m^\tau_{\Lambda_0,a},
\label{eq:A_derivatives}
\end{equation}
and $f^\tau_{\Lambda_0}$ is a concave function of $(\bet,\bu)$, because $\ln Z^\tau_{\Lambda_0}$ is convex in parameters entering $H^\tau_{\Lambda_0}$ linearly.

(P2) \textit{Boundary conditions.}
For a box $\Lambda$ of linear size $\ell$ and any two boundary conditions $\tau,\tau'$ (including periodic ones), the Hamiltonians differ only in the terms of the boundary bonds: at most $2d\ell^{d-1}$ bonds connecting $\Lambda$ with $\partial\Lambda$, and at most $d\ell^{d-1}$ bonds wrapping around the torus. Each such term is bounded by $1$ in magnitude in each Hamiltonian, so that $|H^\tau_\Lambda-H^{\tau'}_\Lambda|\leq2\cdot3d\ell^{d-1}$ for every spin configuration. Hence, for all $\bet$ and $\bu$,
\begin{equation}
|f^\tau_\Lambda-f^{\tau'}_\Lambda|\leq\frac{K_3}{\ell},
\qquad K_3=6d.
\label{eq:A_bc}
\end{equation}

(P3) \textit{Global reversal.}
The global rotation $\theta_i\to\theta_i+\pi$ leaves the coupling terms invariant, since they depend only on angle differences, and maps the system with $(\bet,\bu,\tau)$ to that with $(-\bet,-\bu,\tau+\pi)$. Hence
\begin{equation}
m^\tau_{\Lambda,a}(\bet,\bu)=-m^{\tau+\pi}_{\Lambda,a}(-\bet,-\bu).
\label{eq:A_reversal}
\end{equation}

(P4) \textit{Consistency of Gibbs distributions.}
For a box $\Lambda'\subseteq\Lambda$, the conditional distribution of the spins in $\Lambda'$, given the spins outside $\Lambda'$, is the Gibbs distribution of the system $\Lambda'$ with the boundary condition given by the spins on $\partial\Lambda'$, with the same bonds and fields. This is because the interactions are between nearest neighbors. Consequently, for a quantity $O$ depending only on the spins in $\Lambda'$,
\begin{equation}
\langle O\rangle^\tau_\Lambda=\left\langle\langle O\rangle^{\theta_{\partial\Lambda'}}_{\Lambda'}\right\rangle^\tau_\Lambda ,
\label{eq:A_DLR}
\end{equation}
where $\theta_{\partial\Lambda'}$ denotes the spins on $\partial\Lambda'$. This is often called the DLR property.

All constants $K_1,K_2,\dots$ and $C,C',\dots$ below depend only on $d$ and $h$. In particular, they do not depend on $\beta$, on the bond configuration $B$, on the boundary condition, or on the positions of the boxes. This uniformity is the only property needed to average over $P_{\mathrm{corr}}(B)$ in Eq.~\eqref{eq:Ecorr_decomposition}.

\subsection{Spin-wave bound on the free energy}\label{app:spinwave}

Let $\Lambda_0$ be the whole system, either a box with a boundary condition $\tau$ that does not depend on the field or a torus, and let $\Lambda$ be a box of linear size $\ell$ such that the concentric box $2\Lambda$ of linear size $2\ell$ is contained in $\Lambda_0$. We denote by $\tet$ the field obtained from $\bet$ by reversing its sign inside $\Lambda$: $\tilde{\bm\eta}_i=-\bet_i$ for $i\in\Lambda$ and $\tilde{\bm\eta}_i=\bet_i$ otherwise. We show that
\begin{equation}
\E_\eta\!\left[f(\tet)-f(\bet)\,\middle|\,\bet_{\Lambda},\bet_{\rm out}\right]
\leq\frac{K_1\ell^{d-2}+K_2|\bu|\ell^d}{2|\Lambda_0|},
\label{eq:A_spinwave}
\end{equation}
where $f=f^\tau_{\Lambda_0}$, the conditioning is on the field in $\Lambda$ and outside $2\Lambda$ (denoted by $\bet_{\rm out}$), and $K_1=d\,2^{d+3}\pi^2$ and $K_2=2^{d+2}$.

To derive Eq.~\eqref{eq:A_spinwave}, we introduce a smooth spin-wave profile
\begin{equation}
\alpha_i=\pi\min\left\{1,\frac{2\,\mathrm{dist}(i,W)}{\ell}\right\},
\end{equation}
where $W$ is the set of sites outside $2\Lambda$, including the boundary sites $\partial\Lambda_0$ when $\Lambda_0$ is a box (for a torus, we assume $2\Lambda$ to be a proper subset), and $\mathrm{dist}$ is the $\ell^\infty$ distance. It satisfies $\alpha_i=\pi$ in $\Lambda$, $\alpha_i=0$ outside $2\Lambda$ (in particular on the boundary spins), and $|\alpha_i-\alpha_j|\leq2\pi/\ell$ for nearest neighbors. We define two deformations $R_\pm$ that rotate both the spins and the random fields locally by $\pm\alpha_i$:
\begin{equation}
R_\pm:\ \theta_i\to\theta_i\pm\alpha_i,\qquad \bet_i\to r_{\pm\alpha_i}\bet_i,
\end{equation}
where $r_\alpha$ is the rotation by angle $\alpha$ in the plane. The coupling to the random field is invariant, $(r_\alpha\bet_i)\cdot(r_\alpha\bsig_i)=\bet_i\cdot\bsig_i$. For a bond, with $x=\theta_i-\theta_j-B_{ij}$ and $\Delta=\alpha_i-\alpha_j$,
\begin{align}
-\cos(x+\Delta)-\cos(x-\Delta)
&=-2\cos x+2\cos x\,(1-\cos\Delta)
\nonumber\\
&\leq-2\cos x+\Delta^2 .
\label{eq:A_second_difference}
\end{align}
The first-order terms in $\Delta$ cancel between the two deformations; this is the reason for using the pair $R_\pm$, as in the Mermin--Wagner argument~\cite{MerminWagner1966}. Only the bonds touching $2\Lambda$, at most $2d(2\ell)^d$ in number, have $\Delta\neq0$, and $\Delta^2\leq(2\pi/\ell)^2$ for them. For the uniform field, $-\bu\cdot(r_{\alpha_i}+r_{-\alpha_i})\bsig_i=-2\cos\alpha_i\,\bu\cdot\bsig_i\leq-2\bu\cdot\bsig_i+4|\bu|$ for $i\in2\Lambda$. Writing $H(\theta;\bet)$ for $H^\tau_{\Lambda_0}$, we thus obtain
\begin{align}
&H(R_+\theta;R_+\bet)+H(R_-\theta;R_-\bet)
\nonumber\\
&\leq2H(\theta;\bet)+K_1\ell^{d-2}+K_2|\bu|\ell^d
\label{eq:A_H_sum}
\end{align}
for every spin configuration. We emphasize that Eq.~\eqref{eq:A_second_difference} holds with a constant independent of $B_{ij}$, because the coupling $-\cos(\theta_i-\theta_j-B_{ij})$ depends only on the angle difference.

Next, $Z(R_\pm\bet)=\int d\nu(\theta)\,e^{-\beta H(R_\pm\theta;R_\pm\bet)}$, by the change of integration variables $\theta\to R_\pm\theta$, which leaves the uniform measure invariant and does not affect the boundary spins. Using Eq.~\eqref{eq:A_H_sum} and the Cauchy--Schwarz inequality,
\begin{align}
Z(\bet)
&\leq e^{\beta c/2}\int d\nu(\theta)\,
e^{-\beta[H(R_+\theta;R_+\bet)+H(R_-\theta;R_-\bet)]/2}
\nonumber\\
&\leq e^{\beta c/2}\left[Z(R_+\bet)\,Z(R_-\bet)\right]^{1/2},
\end{align}
with $c=K_1\ell^{d-2}+K_2|\bu|\ell^d$, that is,
\begin{equation}
\frac{f(R_+\bet)+f(R_-\bet)}{2}-f(\bet)\leq\frac{K_1\ell^{d-2}+K_2|\bu|\ell^d}{2|\Lambda_0|}.
\label{eq:A_CS}
\end{equation}
Finally, conditional on $(\bet_\Lambda,\bet_{\rm out})$, the fields $R_\pm\bet$ have the same distribution as $\tet$: inside $\Lambda$, $r_{\pm\pi}\bet_i=-\bet_i$; outside $2\Lambda$, the field is unchanged; and in the annulus $2\Lambda\setminus\Lambda$, the rotated fields $r_{\pm\alpha_i}\bet_i$ are again independent standard Gaussian vectors, independent of the conditioning variables, by the rotational invariance of the Gaussian distribution. Taking the conditional average of Eq.~\eqref{eq:A_CS} gives Eq.~\eqref{eq:A_spinwave}.

Physically, Eq.~\eqref{eq:A_spinwave} states that reversing the random field in a region of linear size $\ell$ changes the total free energy $|\Lambda_0|f$ by at most $O(\ell^{d-2})$ on average, because the spins can follow the reversal by a smooth spin wave. This is the spin-wave part of the Imry--Ma argument.

\subsection{Box magnetization for a fixed boundary condition}\label{app:fixedbc}

With $\Lambda_0$, $\Lambda$, and $\tau$ as in Sec.~\ref{app:spinwave}, we show that
\begin{equation}
\left|\E_\eta\!\left[m^\tau_{\Lambda,a}\right]\right|
\leq\frac{K_1\ell^{d/2-2}+K_2|\bu|\ell^{d/2}}{2\sqrt{2\pi}\,h}.
\label{eq:A_fixedbc}
\end{equation}
For $d<4$ and $|\bu|\leq\ell^{-2}$, the right-hand side decays as $\ell^{-(4-d)/2}$. It is the ratio of the spin-wave cost $\ell^{d-2}$ to the random-field energy $h\ell^{d/2}$ of the Imry--Ma argument. This is the analogue of Eq.~(8.22) of DHP.

Fix the component $a$ and write $\heta=\heta_{\Lambda,a}$, a Gaussian variable with variance $\ell^{-d}$. Let
\begin{equation}
\Phi(t)=\E_\eta\!\left[f(\bet)\,\middle|\,\heta=t\right],\quad
m(t)=\E_\eta\!\left[m^\tau_{\Lambda,a}\,\middle|\,\heta=t\right].
\end{equation}
Conditional on $\heta=t$, the field $\tet$ has $\heta(\tet)=-t$, and the remaining field variables of $\tet$ have the same distribution as those of $\bet$, since they are independent of $\heta$ and symmetric. Hence $\E_\eta[f(\tet)\,|\,\heta=t]=\Phi(-t)$, and Eq.~\eqref{eq:A_spinwave}, averaged further over the variables other than $\heta$, gives $\Phi(-t)-\Phi(t)\leq\epsilon$ for all $t$, with $\epsilon$ the right-hand side of Eq.~\eqref{eq:A_spinwave}. Replacing $t$ by $-t$, we find that the odd function
\begin{equation}
G(t)=\Phi(-t)-\Phi(t)
\quad\text{satisfies}\quad |G(t)|\leq\epsilon .
\end{equation}
By Eq.~\eqref{eq:A_derivatives}, $\Phi'(t)=-(h\ell^d/|\Lambda_0|)\,m(t)$, so that
\begin{equation}
G'(t)=\frac{h\ell^d}{|\Lambda_0|}\left[m(t)+m(-t)\right].
\end{equation}
Since the distribution of $\heta$ is symmetric,
\begin{equation}
\E_\eta[m^\tau_{\Lambda,a}]
=\frac12\E_\eta[m(\heta)+m(-\heta)]
=\frac{|\Lambda_0|}{2h\ell^d}\,\E_\eta[G'(\heta)].
\end{equation}
Gaussian integration by parts gives $\E_\eta[G'(\heta)]=\ell^{d}\,\E_\eta[\heta\,G(\heta)]$, and hence $|\E_\eta[G'(\heta)]|\leq\epsilon\,\ell^{d}\,\E_\eta|\heta|=\epsilon\sqrt{2/\pi}\,\ell^{d/2}$. Substituting $\epsilon$ yields Eq.~\eqref{eq:A_fixedbc}.

The bound~\eqref{eq:A_fixedbc} is stated for a boundary condition that does not depend on the field, since the symmetry of the field distribution has been used. Boundary conditions depending on the field, which arise in Sec.~\ref{app:secondmoment}, are treated in Sec.~\ref{app:uniform}.

\subsection{Uniformity in the boundary condition}\label{app:uniform}

We now set $\bu=0$ in the quantities to be estimated and consider a box $\Lambda$ of linear size $\ell$ as a system by itself. Let
\begin{equation}
X_a(\Lambda)=\sup_\tau m^\tau_{\Lambda,a},
\qquad
Y_a(\Lambda)=\inf_\tau m^\tau_{\Lambda,a},
\end{equation}
where the supremum and infimum are over all boundary conditions $\tau$ on $\partial\Lambda$; $X_a(\Lambda)$ and $Y_a(\Lambda)$ depend on the field in $\Lambda$ only. We show that, for $d\leq3$,
\begin{equation}
\E_\eta\!\left[\sup_\tau\left|m^\tau_{\Lambda,a}\right|\right]
\leq C\,\ell^{-(4-d)/[2(8-d)]},
\label{eq:A_uniform}
\end{equation}
which is Eq.~(3.5) of DHP. The derivation proceeds in five steps.

\textit{(a) Comparison with periodic boundary conditions.} Let $\Lambda^{\rm per}$ be the torus obtained from $\Lambda$ by identifying its opposite faces, with arbitrary values of the bonds that wrap around. By Eq.~\eqref{eq:A_bc}, $|f^\tau_\Lambda-f^{\rm per}_\Lambda|\leq K_3/\ell$ for all $\tau$, $\bet$, and $\bu$.

\textit{(b) Periodic system.} We partition the torus into boxes $\Lambda'$ of linear size $\ell'\leq\ell/4$, so that $2\Lambda'$ is a proper subset of the torus, and apply Eq.~\eqref{eq:A_fixedbc} to each of them with $\Lambda_0=\Lambda^{\rm per}$. This replaces the use of translation invariance in DHP [their Eq.~(8.30)], which does not hold for a fixed bond configuration. Averaging over the boxes and choosing $\ell'$ of order $\min(|\bu|^{-1/2},\ell/4)$ for $|\bu|\leq1$ (only such values are used below), we obtain
\begin{equation}
\left|\E_\eta\!\left[m^{\rm per}_{\Lambda,a}(\bu)\right]\right|\leq C\left(|\bu|\vee\ell^{-2}\right)^{1-d/4},
\label{eq:A_per}
\end{equation}
where $x\vee y=\max(x,y)$. For $\ell$ smaller than a fixed constant, Eq.~\eqref{eq:A_per} holds trivially with a suitable $C$, since $|m^{\rm per}_{\Lambda,a}|\leq1$.

\textit{(c) Integration over the auxiliary field.} By Eq.~\eqref{eq:A_derivatives}, $F(\bu):=\E_\eta[f^{\rm per}_\Lambda(\bet,\bu)]$ satisfies
\begin{equation}
|F(\bu)-F(0)|\leq|\bu|\max_{0\leq t\leq1}\left|\E_\eta[\bm m^{\rm per}_{\Lambda}(t\bu)]\right|
\leq C\left(|\bu|\vee\ell^{-2}\right)^{2-d/4},
\label{eq:A_psi}
\end{equation}
where $\bm m^{\rm per}_\Lambda=(m^{\rm per}_{\Lambda,x},m^{\rm per}_{\Lambda,y})$.

\textit{(d) Convexity.} Let $\tau^*(\bet)$ be a boundary condition that attains $X_a(\Lambda)$ at $\bu=0$; it exists because $m^\tau_{\Lambda,a}$ is continuous in $\tau$, which takes values in a compact set. The function $g(\bu)=-\E_\eta[f^{\tau^*(\bet)}_\Lambda(\bet,\bu)]$ is convex, and by Eq.~\eqref{eq:A_derivatives}, $\partial g/\partial u_a$ at $\bu=0$ equals $\E_\eta[X_a(\Lambda)]$. For $s>0$, convexity gives $\partial_{u_a}g(0)\leq[g(s\bm e_a)-g(0)]/s$, where $\bm e_a$ is the unit vector in direction $a$. By steps (a) and (c),
\begin{equation}
\E_\eta[X_a(\Lambda)]\leq\frac{C\left(s\vee\ell^{-2}\right)^{2-d/4}+2K_3/\ell}{s}.
\end{equation}
The choice $s=\ell^{-4/(8-d)}$ balances the two terms and gives $\E_\eta[X_a(\Lambda)]\leq\varepsilon_1(\ell):=C\ell^{-\kappa_d}$ with $\kappa_d=(4-d)/(8-d)$. By the reversal property~\eqref{eq:A_reversal} and the symmetry of the field distribution, $\E_\eta[Y_a(\Lambda)]=-\E_\eta[X_a(\Lambda)]$. Since $X_a\geq Y_a$, it follows that $0\leq\E_\eta[X_a(\Lambda)]\leq\varepsilon_1(\ell)$ and $-\varepsilon_1(\ell)\leq\E_\eta[Y_a(\Lambda)]\leq0$.

\textit{(e) Fluctuations.} Step (d) bounds the averages of $X_a$ and $Y_a$, whereas Eq.~\eqref{eq:A_uniform} requires $\sup_\tau|m^\tau_{\Lambda,a}|=\max(X_a,-Y_a)$, which could be large if $X_a$ and $Y_a$ fluctuated strongly with the field. To exclude this, we partition $\Lambda$ into $N'$ boxes $\Lambda_j$ of linear size $\ell''$. By the consistency property~\eqref{eq:A_DLR}, for every $\tau$,
\begin{equation}
m^\tau_{\Lambda,a}=\frac{1}{N'}\sum_{j=1}^{N'}\left\langle m^{\theta_{\partial\Lambda_j}}_{\Lambda_j,a}\right\rangle^\tau_\Lambda\in\left[\bar Y,\bar X\right],
\end{equation}
where $\bar X=N'^{-1}\sum_jX_a(\Lambda_j)$ and $\bar Y=N'^{-1}\sum_jY_a(\Lambda_j)$. Thus $\sup_\tau|m^\tau_{\Lambda,a}|\leq|\bar X|+|\bar Y|$. The variables $X_a(\Lambda_j)$ for different $j$ depend on the fields in disjoint boxes and are therefore independent; they take values in $[-1,1]$ and have averages in $[0,\varepsilon_1(\ell'')]$. Hence $\E_\eta|\bar X|\leq\varepsilon_1(\ell'')+N'^{-1/2}$, and similarly for $\bar Y$. Choosing $\ell''=\ell^{1/2}$, so that $N'=\ell^{d/2}$,
\begin{equation}
\E_\eta\!\left[\sup_\tau|m^\tau_{\Lambda,a}|\right]\leq2C\ell^{-\kappa_d/2}+2\ell^{-d/4},
\end{equation}
which gives Eq.~\eqref{eq:A_uniform}, since $\kappa_d/2<d/4$.

\subsection{Four dimensions}\label{app:d4}

In $d=4$, the bound in Eq.~(A16) does not decay
with increasing box size, reflecting the marginal
nature of the Imry--Ma argument.
To overcome this difficulty, we follow the multiscale
approach of DHP.
The key idea is to identify boxes in which the random
field suppresses the magnetization at different length
scales and show that these boxes cover almost the
entire system.

The proof proceeds as follows.
First, using the spin-wave bound \eqref{eq:A_spinwave}, we establish
the single-box estimate \eqref{eq:A_goodbox}, which shows that
a sufficiently negative spatial average of a random-field
component bounds the corresponding conditional
expectation of the magnetization from above.
Second, we construct a hierarchy of nested boxes
and select the largest boxes satisfying this condition,
called good boxes.
We show that the probability of a site remaining
uncovered is small, as quantified in
Eq.~\eqref{eq:uncovered-sites}.
Third, we bound the contribution from the selected
good boxes.
Since the selection of a box also depends on its
ancestors, we introduce an approximating event
determined by the variables appearing in the
conditional estimate \eqref{eq:A_goodbox}.
This yields Eq.~\eqref{eq:A_beta}.
Combining the bounds on uncovered sites and selected
good boxes gives Eq.~\eqref{eq:A_d4_fixed} for a fixed boundary
condition independent of the random field.
Finally, using the convexity and fluctuation arguments
of Sec.~A~4, we extend the estimate to arbitrary
boundary conditions and obtain Eq.~\eqref{eq:A_uniform_d4}.
All bounds are uniform in the fixed bond configuration $B$.

\textit{Good boxes.} Let $d=4$, let $\Lambda_0$, $\Lambda$ (of linear size $\ell$), and $\tau$ be as in Sec.~\ref{app:spinwave}, and let $|\bu|\leq\ell^{-2}$. Fix $a$, write $\heta=\heta_{\Lambda,a}$, and define $m(t)=\E_\eta[m^\tau_{\Lambda,a}\,|\,\heta=t,\bet_{\rm out}]$, the conditional average for fixed field $\bet_{\rm out}$ outside $2\Lambda$. Then, for $0<\delta\leq1$,
\begin{equation}
m(t)\leq\delta\quad\text{whenever}\quad \ell^2t\leq t_\delta ,
\label{eq:A_goodbox}
\end{equation}
for every $\bet_{\rm out}$, where $t_\delta<0$ is defined by $\int_{-\infty}^{t_\delta}dz\,e^{-z^2/2}/\sqrt{2\pi}=p_\delta$ with
\begin{equation}
p_\delta=\frac{K_4}{6\sqrt{2\pi}}\,e^{-2K_4^2/\delta^2},\qquad K_4=\frac{K_1+K_2}{2h}.
\label{eq:A_pdelta}
\end{equation}
Note that $p_\delta<1/2$, since $K_4e^{-2K_4^2/\delta^2}\leq(\delta/2)e^{-1/2}$, so that $t_\delta<0$ is well defined; that $p_\delta\geq e^{-C/\delta^2}$ for $\delta\leq1$; and that $\ell^2\heta$ is a standard Gaussian variable in $d=4$. In words, if the field averaged over $\Lambda$ points against the direction $a$ by a sufficiently large multiple of its typical fluctuation, then the conditional average of the magnetization of $\Lambda$ in direction $a$ is at most $\delta$. This is a variant of Lemma~8.6 of DHP.

The derivation is as follows. Let $\Phi(t)=\E_\eta[f\,|\,\heta=t,\bet_{\rm out}]$. As in Sec.~\ref{app:fixedbc}, Eq.~\eqref{eq:A_spinwave} gives $|\Phi(t)-\Phi(-t)|\leq\epsilon:=(K_1+K_2)\ell^2/(2|\Lambda_0|)$, and $\Phi'(t)=-(h\ell^4/|\Lambda_0|)\,m(t)$. Since $f$ is concave in the field, so is $\Phi$, and $m(t)$ is therefore a nondecreasing function of $t$. Let $w(t)=m(t)+m(-t)\in[-2,2]$. Then
\begin{equation}
\int_0^Tw(t)\,dt=\frac{|\Lambda_0|}{h\ell^4}\left[\Phi(-T)-\Phi(T)\right]\leq\frac{K_4}{\ell^2},
\end{equation}
for all $T>0$. Let $I$ be the set of $t\geq0$ with $w(t)\leq\delta$, and $I_T=I\cap[0,T]$. Since $w>\delta$ outside $I$ and $w\geq-2$ on $I$, we have $\int_0^Tw\geq\delta(T-|I_T|)-2|I_T|$, and hence $|I_T|\geq(\delta T-K_4/\ell^2)/3$. Taking $T_0=2K_4/(\delta\ell^2)$, we find $|I_{T_0}|\geq K_4/(3\ell^2)$, and, since the probability density of $\heta$ on $[0,T_0]$ is at least $(\ell^2/\sqrt{2\pi})e^{-2K_4^2/\delta^2}$,
\begin{equation}
\mathrm{Prob}\left[w(\heta)\leq\delta\right]\geq\frac{K_4}{3\sqrt{2\pi}}\,e^{-2K_4^2/\delta^2}.
\end{equation}
If $w(t)\leq\delta$, then $m(t)\leq\delta$ or $m(-t)\leq\delta$. By the symmetry of the distribution of $\heta$, $\mathrm{Prob}[m(\heta)\leq\delta]\geq p_\delta$, with $p_\delta$ given by Eq.~\eqref{eq:A_pdelta}. Since $m$ is nondecreasing, the set of $t$ with $m(t)\leq\delta$ is a half-line extending to $-\infty$ with Gaussian probability at least $p_\delta$, and it therefore contains all $t$ with $\ell^2t\leq t_\delta$. This proves Eq.~\eqref{eq:A_goodbox}.

\textit{Multiscale covering.} Let $\Lambda_*$ be a box of linear size $L$ with $2\Lambda_*\subseteq\Lambda_0$, $\tau$ independent of the field, and $|\bu|\leq L^{-2}$. (Here $\Lambda_*$ is a box inside the system $\Lambda_0$, to be distinguished from the torus $\Lambda_L$ of the main text.) Following Lemma~8.7 of DHP, one chooses $\delta=C_0/\sqrt{\ln\ln L}$ with a large constant $C_0$ (we consider $L$ large enough that $\delta\leq1/2$; smaller $L$ are accounted for by adjusting the constants), an odd integer $k$ of order $\exp[(\ln L)^{1/4}]$, and $l_{\max}$ levels, where $l_{\max}=\lfloor\ln L/(2\ln k)\rfloor$ is the largest integer with $k^{l_{\max}}\leq\sqrt L$, so that $l_{\max}$ is of order $(\ln L)^{3/4}$. For $l=0,1,\dots,l_{\max}$, the box $\Lambda_*$ is partitioned into $k^{4l}$ boxes of linear size $L/k^l$, which are nested across levels (we first take $L$ to be a multiple of $k^{l_{\max}}$ and treat general $L$ below; see also the construction of DHP, their Eq.~(7.23)).

A box $\Lambda'$ of linear size $\ell'$ is called good if $\ell'^2\heta_{\Lambda',a}\leq t_\delta$, and $\mathsf G$ denotes the set of good boxes that are not contained in a larger good box. The boxes in $\mathsf G$ are disjoint, and every box $\Lambda'$ of the hierarchy satisfies $2\Lambda'\subseteq\Lambda_0$ and $|\bu|\leq\ell'^{-2}$, so that Eq.~\eqref{eq:A_goodbox} applies to it. The following two estimates correspond to Eqs.~(8.52) and (8.53) of DHP; we give the essential steps.
\begin{enumerate}
\item[($\alpha$)] \textit{Uncovered sites.} 
A site $v\in\Lambda_*$ is not covered by $\mathsf G$ with probability at most $\exp(-c\sqrt{\ln L})$. To see this, let $D_0\supset D_1\supset\dots\supset D_{l_{\max}}$ be the boxes of the hierarchy containing $v$, with linear sizes $\ell_l=L/k^l$, and let $Z_l=\ell_l^2\heta_{D_l,a}$, a standard Gaussian variable in $d=4$. Then
\begin{equation}
Z_l=S_l+k^{-2}Z_{l+1},\qquad S_l=\ell_l^{-2}\sum_{i\in D_l\setminus D_{l+1}}\eta_{i,a},
\end{equation}
for $l=0,1,\dots,l_{\max}-1$, where the $S_l$ are independent Gaussian variables with variance $1-k^{-4}$. The site $v$ is uncovered if $Z_l>t_\delta$ for all $l$. Unless $|Z_{l+1}|>k$ for some $l$, an event of probability at most $2l_{\max}e^{-k^2/2}$, this requires $S_l>t_\delta-k^{-1}$ for all $l$, an intersection of independent events each of probability at most $1-p_\delta/2$ for large $L$. Indeed, $\mathrm{Prob}[S_l\leq t_\delta-k^{-1}]/p_\delta\to1$ as $L\to\infty$, because $|t_\delta|\leq C/\delta$ grows only as $\sqrt{\ln\ln L}$, whereas the shift $k^{-1}$ of the threshold and the change $k^{-4}$ of the variance decrease faster than any power of $\ln L$. Hence the probability is at most $(1-p_\delta/2)^{l_{\max}}+2l_{\max}e^{-k^2/2}$. 
Since
$p_\delta\geq e^{-C/\delta^2}
=(\ln L)^{-C/C_0^2}$
and $l_{\max}$ is of order $(\ln L)^{3/4}$,
we obtain, for sufficiently large $C_0$,
\begin{equation}
\begin{aligned}
\mathrm{Prob}\left(
v\notin\bigcup_{D\in\mathsf G}D
\right)
&\leq
\left(1-\frac{p_\delta}{2}\right)^{l_{\max}}
+2l_{\max}e^{-k^2/2}\\
&\leq
\exp\left(-c\sqrt{\ln L}\right),
\label{eq:uncovered-sites}
\end{aligned}
\end{equation}
for all sufficiently large $L$,
with a constant $c>0$ independent of $v$,
$B$, and $\beta$.

\item[($\beta$)] \textit{Good boxes in the covering.} For every box $D$ of the hierarchy, of linear size $\ell$,
\begin{equation}
\E_\eta\!\left[\mathbf 1_{\{D\in\mathsf G\}}m^\tau_{D,a}\right]\leq\delta\,\mathrm{Prob}[D\in\mathsf G]+\frac{C}{(\ln L)^2}.
\label{eq:A_beta}
\end{equation}
The event $\{D\in\mathsf G\}$ means that $D$ is good and that all its ancestors $A_j$, the larger boxes of the hierarchy containing $D$, with linear sizes $\ell_j\geq k\ell$, are bad. It is not determined by $\heta_{D,a}$ and the field outside $2D$ alone, because $\heta_{A_j,a}$ includes the field in $2D$. We therefore write $\heta_{A_j,a}=T_j+U_j$, with
\begin{equation}
T_j=\frac{1}{|A_j|}\sum_{i\in A_j\setminus2D}\eta_{i,a},\qquad
U_j=\frac{1}{|A_j|}\sum_{i\in A_j\cap2D}\eta_{i,a},
\end{equation}
and introduce the event
\begin{equation}
E_D=\{\ell^2\heta_{D,a}\leq t_\delta\}\cap\bigcap_j\{\ell_j^2T_j>t_\delta\},
\end{equation}
which is determined by $\heta_{D,a}$ and the field outside $2D$. With $a_L=(\ln L)^{-3}$, the events $\{D\in\mathsf G\}$ and $E_D$ can differ only if, for some $j$, either $\ell_j^2|U_j|>a_L$ or $|\ell_j^2T_j-t_\delta|\leq a_L$. Since $\ell_j^2U_j$ is Gaussian with variance at most $16(\ell/\ell_j)^4\leq16k^{-4}$ and $\ell_j^2T_j$ is Gaussian with variance at least $1/2$, these events have probabilities at most $2\exp(-k^4a_L^2/32)$ and $2a_L/\sqrt\pi$, respectively. 
Summing over the at most $l_{\max}$ ancestors, the probability that $\{D\in\mathsf G\}$ and $E_D$ differ is at most $C/(\ln L)^2$. On $E_D$, Eq.~\eqref{eq:A_goodbox} gives $\E_\eta[m^\tau_{D,a}\,|\,\heta_{D,a},\bet_{\rm out}]\leq\delta$; together with $|m^\tau_{D,a}|\leq1$, this yields Eq.~\eqref{eq:A_beta}.
\end{enumerate}

Decomposing $\Lambda_*$ into the boxes of $\mathsf G$ and the uncovered sites, $m^\tau_{\Lambda_*,a}=\sum_{D}(|D|/|\Lambda_*|)\mathbf 1_{\{D\in\mathsf G\}}m^\tau_{D,a}+|\Lambda_*|^{-1}\sum_{v\ \text{uncovered}}\langle\sigma_{v,a}\rangle^\tau_{\Lambda_0}$, where the sum runs over all boxes of the hierarchy. Since the boxes of $\mathsf G$ are disjoint, $\sum_D(|D|/|\Lambda_*|)\,\mathrm{Prob}[D\in\mathsf G]\leq1$, while $\sum_D|D|/|\Lambda_*|=l_{\max}+1$. Hence Eq.~\eqref{eq:A_beta} and estimate ($\alpha$) give
\begin{equation}
\E_\eta[m^\tau_{\Lambda_*,a}]\leq\delta+\frac{C\,(l_{\max}+1)}{(\ln L)^2}+e^{-c\sqrt{\ln L}}\leq\frac{C}{\sqrt{\ln\ln L}},
\label{eq:A_d4_fixed}
\end{equation}
and the same bound for $-\E_\eta[m^\tau_{\Lambda_*,a}]$ by the reversal property~\eqref{eq:A_reversal}.

We stress that the estimates ($\alpha$) and ($\beta$) concern the Gaussian field only. The interaction enters solely through Eq.~\eqref{eq:A_goodbox} and the bound $|m^\tau_{D,a}|\leq1$, and the constants depend neither on $B$ nor on $\beta$. The construction of DHP therefore applies to the present model with an arbitrary fixed bond configuration.

\textit{General linear sizes.} If $L$ is not a multiple of $k^{l_{\max}}$, each coordinate interval of $\Lambda_*$ is divided into $k$ consecutive intervals whose lengths differ by at most one, and the division is repeated at each level. The side lengths of the boxes at level $l$ then differ from $L/k^l$ by less than $2$, and hence by a factor $1+O(L^{-1/2})$, since $L/k^l\geq\sqrt L$. Equation~\eqref{eq:A_spinwave} holds for a box whose side lengths lie between $\ell$ and $2\ell$, with $\ell$ the smallest side length, if $K_1$ and $K_2$ are replaced by larger constants depending only on $d$, and the derivation of Eq.~\eqref{eq:A_goodbox} goes through with $\ell^2\heta_{\Lambda,a}$ replaced by $|\Lambda|^{1/2}\heta_{\Lambda,a}$, which is again a standard Gaussian variable, and with a modified $K_4$. In the definition of good boxes and in ($\alpha$) and ($\beta$), $\ell'^2\heta_{\Lambda',a}$ is replaced accordingly by $|\Lambda'|^{1/2}\heta_{\Lambda',a}$. The coefficient $k^{-2}$ in the recursion for $Z_l$ becomes $(|D_{l+1}|/|D_l|)^{1/2}$, the variance of $S_l$ becomes $1-|D_{l+1}|/|D_l|$, and the bounds on the variances in ($\beta$) change by factors $1+O(L^{-1/2})$. The estimates ($\alpha$) and ($\beta$), and hence Eq.~\eqref{eq:A_d4_fixed}, therefore hold with modified constants depending only on $d$ and $h$.

\textit{Uniformity in the boundary condition.} Steps (a)--(e) of Sec.~\ref{app:uniform} apply with Eq.~\eqref{eq:A_per} replaced by $|\E_\eta[m^{\rm per}_{\Lambda,a}(\bu)]|\leq C/\sqrt{\ln\ln(\ell\wedge|\bu|^{-1})}$, with $x\wedge y=\min(x,y)$, for $\ell$ and $|\bu|^{-1}$ larger than a fixed constant (otherwise the trivial bound $1$ is used). This follows from Eq.~\eqref{eq:A_d4_fixed} applied to the boxes of a partition of the torus into boxes of linear size of order $\min(\ell/4,|\bu|^{-1/2})$; the logarithm of this size is of the same order as $\ln(\ell\wedge|\bu|^{-1})$. In step (d), one takes $s=\ell^{-1/2}$. The result is
\begin{equation}
\E_\eta\!\left[\sup_\tau|m^\tau_{\Lambda,a}|\right]\leq\frac{C}{\sqrt{\ln\ln\ell}}\qquad(d=4),
\label{eq:A_uniform_d4}
\end{equation}
which is Eq.~(3.6) of DHP at zero uniform field.

\subsection{From box magnetizations to the second moment}\label{app:secondmoment}

The estimates~\eqref{eq:A_uniform} and \eqref{eq:A_uniform_d4} control the thermal average of the magnetization of a box, uniformly in the boundary condition. The distribution function $P_{\mathrm{corr}}$, however, requires the second moment $\langle|M_{\Lambda_L}|^2\rangle$. The two quantities differ in general. For example, the Ising ferromagnet below the critical temperature with free boundary conditions has $\langle M\rangle=0$, whereas $\langle|M|\rangle$ converges to the spontaneous magnetization in the thermodynamic limit. Moreover, gauge symmetry alone does not relate $\langle|M|^2\rangle$ to $|\langle M\rangle|^2$. We note also that the estimates concern spatial averages over boxes, not local magnetizations $\langle\bsig_i\rangle$. The argument below uses box averages only.

Consider the torus $\Lambda_L$ with periodic boundary conditions and fixed $(B,\bet)$, and partition it into $N_s=(L/\ell)^d$ boxes $\tilde\Lambda_j$ of linear size $\ell$. Since $M_{\Lambda_L}=N_s^{-1}\sum_jM_{\tilde\Lambda_j}$, where $M_{\tilde\Lambda_j}$ is the magnetization~\eqref{eq:QM} of $\tilde\Lambda_j$,
\begin{equation}
\left\langle|M_{\Lambda_L}|^2\right\rangle
=\frac{1}{N_s^2}\sum_{j,j'}\left\langle\mathrm{Re}\left(M_{\tilde\Lambda_j}\overline{M_{\tilde\Lambda_{j'}}}\right)\right\rangle .
\end{equation}
The diagonal terms contribute at most $N_s^{-1}$, since $|M_{\tilde\Lambda_j}|\leq1$. For $j\neq j'$, $M_{\tilde\Lambda_{j'}}$ depends only on the spins outside $\tilde\Lambda_j$, and the consistency property~\eqref{eq:A_DLR} gives
\begin{equation}
\left\langle M_{\tilde\Lambda_j}\overline{M_{\tilde\Lambda_{j'}}}\right\rangle
=\left\langle\langle M_{\tilde\Lambda_j}\rangle^{\theta_{\partial\tilde\Lambda_j}}_{\tilde\Lambda_j}\overline{M_{\tilde\Lambda_{j'}}}\right\rangle,
\end{equation}
whose real part is bounded by $\sup_\tau|\langle M_{\tilde\Lambda_j}\rangle^\tau_{\tilde\Lambda_j}|$. Here $\langle\cdot\rangle^\tau_{\tilde\Lambda_j}$ is the Gibbs distribution of $\tilde\Lambda_j$ as a system by itself, with the bonds inside $\tilde\Lambda_j$ and those joining it to $\partial\tilde\Lambda_j$, the fields in $\tilde\Lambda_j$, and the boundary condition $\tau$. Summing over $j'$, we obtain
\begin{equation}
\left\langle|M_{\Lambda_L}|^2\right\rangle
\leq\frac{1}{N_s}+\frac{1}{N_s}\sum_{j=1}^{N_s}\sup_\tau\left|\langle M_{\tilde\Lambda_j}\rangle^\tau_{\tilde\Lambda_j}\right| .
\label{eq:A_second_moment}
\end{equation}
The supremum includes the boundary conditions $\theta_{\partial\tilde\Lambda_j}$, which depend on the field and on the thermal configuration outside $\tilde\Lambda_j$; this is why the uniform estimates of Secs.~\ref{app:uniform} and \ref{app:d4} are needed.

If $L$ is not a multiple of $\ell$, the boxes have side lengths between $\ell$ and $2\ell$ and different volumes. With $w_j=|\tilde\Lambda_j|/|\Lambda_L|$, we have $M_{\Lambda_L}=\sum_jw_jM_{\tilde\Lambda_j}$ and $\sum_jw_j=1$, and the same argument gives Eq.~\eqref{eq:A_second_moment} with $1/N_s$ in the first term replaced by $\sum_jw_j^2\leq\max_jw_j\leq(2\ell/L)^d$ and with $1/N_s$ in the sum replaced by $w_j$. The estimates below then change only by constants.

Taking $\E_\eta$ of Eq.~\eqref{eq:A_second_moment}, using $|\langle M\rangle^\tau|\leq|m^\tau_x|+|m^\tau_y|$ and Eqs.~\eqref{eq:A_uniform} and \eqref{eq:A_uniform_d4}, and choosing $\ell=\sqrt L$, we obtain
\begin{equation}
\E_\eta\!\left[\left\langle|M_{\Lambda_L}|^2\right\rangle\right]
\leq L^{-d/2}+
\begin{cases}
C\,L^{-(4-d)/[4(8-d)]} & (d\leq3),\\[2pt]
C/\sqrt{\ln\ln\sqrt L} & (d=4).
\end{cases}
\end{equation}
This proves Eq.~\eqref{eq:theorem_second_moment}. The particular choice $\ell=\sqrt L$ is not essential for the qualitative conclusion: taking $L\to\infty$ at fixed $\ell$ in Eq.~\eqref{eq:A_second_moment} and then $\ell\to\infty$ also gives $\E_\eta[\langle|M_{\Lambda_L}|^2\rangle]\to0$. Equation~\eqref{eq:theorem_sup} is Eq.~\eqref{eq:A_uniform} or \eqref{eq:A_uniform_d4} with $\ell=L$, combined with $|\langle M\rangle^\tau|\leq|m^\tau_x|+|m^\tau_y|$. Since all constants are independent of $\beta$ and $B$, the proof of Theorem~\ref{thm:main} is complete.

\subsection{Technical details and relation to DHP}\label{app:remarks}

For the convenience of readers who wish to check the argument against DHP or to write it at the level of mathematical rigor, we list the correspondence and the points that have been treated briefly.

\textit{Assumptions of DHP.} (i) DHP assume an interaction $\Psi(\bsig_i,\bsig_j)$ that is twice continuously differentiable, independent of the bond, and invariant under $O(n)$ [their Eq.~(3.1)]. The present interaction $-\cos(\theta_i-\theta_j-B_{ij})$ depends on the bond and is invariant only under $SO(2)$. The proof uses only the invariance under the one-parameter group of rotations and the uniform bound~\eqref{eq:A_second_difference} on the second difference, both satisfied with constants independent of $B$; the global reversal used in Eq.~\eqref{eq:A_reversal} is the rotation by $\pi$, which belongs to $SO(2)$. (ii) DHP use translation invariance of the periodic system [their Eqs.~(6.12) and (8.30), and the beginning of the proof of their Theorem~5]. This is replaced by the partition of the torus in step (b) of Sec.~\ref{app:uniform}. (iii) The bound on the second moment, Sec.~\ref{app:secondmoment}, is not part of DHP.

\textit{Simplifications.} Since $n=2$ and the rotation by $\pi$ reverses both components of the field, we condition on the full field in $\Lambda$ in Eq.~\eqref{eq:A_spinwave}, whereas DHP condition on one component for general $n$ [their Proposition~8.4]. The Gaussian integration by parts in Sec.~\ref{app:fixedbc} replaces their Proposition~8.1, and the measure estimate in Sec.~\ref{app:d4} replaces their Proposition~8.2. In step (e) of Sec.~\ref{app:uniform}, a variance bound replaces the Hoeffding inequality; DHP obtain stronger probability bounds, which are not needed here.

The exponents obtained are not optimal. For instance, with $\gamma_d=(4-d)/[2(8-d)]$, the choice $\ell=L^{d/(d+\gamma_d)}$ in Sec.~\ref{app:secondmoment} instead of $\ell=\sqrt L$ balances the two terms on the right-hand side of Eq.~\eqref{eq:A_second_moment} after the average over the field, and improves the exponent in Eq.~\eqref{eq:theorem_second_moment} for $d\leq3$ from $\gamma_d/2$ to $d\gamma_d/(d+\gamma_d)$; for example, from $1/20$ to $3/31$ in $d=3$, where $\gamma_3=1/10$.

\textit{Points treated briefly.} (i) Existence and measurability of a maximizing boundary condition $\tau^*(\bet)$ [DHP, Eq.~(8.32)]: $m^\tau_{\Lambda,a}$ is continuous in $\tau$ and in $\bet$, and the set of boundary conditions is compact. (ii) Differentiation under the average over the field: all derivatives in Eq.~\eqref{eq:A_derivatives} are bounded. (iii) Divisibility of $L$ and $\ell$ by the linear sizes of the sub-boxes: one may use boxes whose linear sizes lie between $\ell$ and $2\ell$, which changes only the constants. For the partitions in Secs.~\ref{app:uniform} and \ref{app:secondmoment}, each coordinate interval is divided into intervals of lengths between $\ell$ and $2\ell$, and Eq.~\eqref{eq:A_spinwave} and the estimates derived from it hold with $\ell$ the smallest side length and with constants depending only on $d$ and $h$. The hierarchy of boxes in four dimensions is treated in the paragraph on general linear sizes in Sec.~\ref{app:d4}. (iv) In four dimensions, the estimates ($\alpha$) and ($\beta$) are given above in condensed form; the corresponding detailed treatment of the choices of $k$, $l_{\max}$, and $\delta$ and of the Gaussian estimates is in Eqs.~(8.52)--(8.64) of DHP. Since these steps involve only the Gaussian field and Eq.~\eqref{eq:A_goodbox}, they can be transcribed without change. (v) All estimates are uniform in $\beta$. DHP state zero-temperature versions of their results (their Sec.~3), with free energies replaced by ground-state energies; we have not checked the corresponding versions of the present estimates in detail. (vi) For free boundary conditions on $\Lambda_L$, the boundary conditions of the boxes adjacent to the surface in Sec.~\ref{app:secondmoment} are partly free. The arguments of Secs.~\ref{app:uniform}--\ref{app:secondmoment} apply if the supremum over $\tau$ is extended to such boundary conditions, since Eqs.~\eqref{eq:A_bc} and \eqref{eq:A_reversal} hold for them as well.

%%%%%%%%%%%%%%%%%%%%%%%%%%%%%%%%%%%%%%%%%%%%%%%%%%%%%%%%%%%%%
%%%%%%%%%%%%%%%%%%%%%%%%%%%%%%%%%%%%%%%%%%%%%%%%%%%%%%%%%%%%%%%%%%%%%%%%%%%%
\bibliography{references}

@article{Guerra2003,
  author  = {Guerra, Francesco},
  title   = {{Broken replica symmetry bounds in the mean field spin glass model}},
  journal = {Commun. Math. Phys.},
  volume  = {233},
  pages   = {1},
  year    = {2003},
  doi     = {10.1007/s00220-002-0773-5}
}

@article{Talagrand2006,
author = {Talagrand, Michel},
doi = {10.4007/annals.2006.163.221},
issn = {0003-486X},
journal = {Ann. Math.},
month = {jan},
number = {1},
pages = {221},
title = {{The Parisi formula}},
volume = {163},
year = {2006}
}

@book{Panchenko2013,
  author    = {Panchenko, Dmitry},
  title     = {{The Sherrington--Kirkpatrick Model}},
  publisher = {Springer},
  address   = {New York},
  year      = {2013}
}

@article{Nishimori1980,
  title={Exact results and critical properties of the {Ising} model with competing interactions},
   doi = {10.1088/0022-3719/13/21/012},
  author={Nishimori, Hidetoshi},
  journal={J. Phys. C},
  volume={13},
  number={21},
  pages={4071},
  year={1980},
  publisher={IOP Publishing}
}

@article{Nishimori1981,
  title={Internal energy, specific heat and correlation function of the bond-random {Ising} model},
  author={Nishimori, Hidetoshi},
  journal={Prog. Theor. Phys.},
  volume={66},
  number={4},
  pages={1169},
  year={1981},
  publisher={Oxford University Press},
  doi = {10.1143/PTP.66.1169}
}

@book{Nishimori2001,
  title={{Statistical Physics of Spin Glasses and Information Processing: An Introduction}},
  author={Nishimori, Hidetoshi},
  year={2001},
  publisher={Oxford University Press},
address = {Oxford}
}

@article{Ozeki1993,
author = {Ozeki, Yukiyasu and Nishimori, Hidetoshi},
journal = {J. Phys. A},
pages = {3399},
title = {{Phase diagram of gauge glasses}},
volume = {26},
doi={10.1088/0305-4470/26/14/009},
year = {1993}
}

@article{Nishimori2022,
doi = {10.1088/1751-8121/ac44ef},
year = {2022},
month = {jan},
publisher = {IOP Publishing},
volume = {55},
number = {4},
pages = {045001},
author = {Hidetoshi Nishimori},
title = {Analyticity of the energy in an {Ising} spin glass with correlated disorder},
journal = {J. Phys. A},
}

@article{Nishimori2024,
author = {Nishimori, Hidetoshi},
doi = {10.1103/PhysRevE.110.064108},
journal = {Phys. Rev. E},
pages = {064108},
publisher = {American Physical Society},
title = {{Anomalous distribution of magnetization in an Ising spin glass with correlated disorder}},
volume = {110},
year = {2024}
}

@article{Nishimori2025,
author = {Nishimori, Hidetoshi},
doi = {10.1103/PhysRevE.111.044109},
journal = {Phys. Rev. E},
pages = {044109},
publisher = {American Physical Society},
title = {{Instability of the ferromagnetic phase under random fields in an Ising spin glass with correlated disorder}},
volume = {111},
year = {2025}
}

@article{Nishimori2025a,
author = {Nishimori, Hidetoshi and Ohzeki, Masayuki and Okuyama, Manaka},
title = {{Temperature chaos as a logical consequence of the reentrant transition in spin glasses}},
journal = {Phys. Rev. E},
volume = {112},
number = {4},
pages = {044140},
year = {2025},
doi = {10.1103/qp1w-qcbs}
}

@article{Reger1991,
  author  = {Reger, J. D. and Tokuyasu, T. A. and Young, A. P. and Fisher, Matthew P. A.},
  title   = {{Vortex-glass transition in three dimensions}},
  journal = {Phys. Rev. B},
  volume  = {44},
  pages   = {7147(R)},
  year    = {1991},
  doi     = {10.1103/PhysRevB.44.7147}
}

@article{Olson2000,
  author  = {Olson, Thomas and Young, A. P.},
  title   = {{Finite temperature ordering in the three-dimensional gauge glass}},
  journal = {Phys. Rev. B},
  volume  = {61},
  pages   = {12467},
  year    = {2000},
  doi     = {10.1103/PhysRevB.61.12467}
}

@article{Katzgraber2005,
  author  = {Katzgraber, Helmut G. and Campbell, I. A.},
  title   = {{Dynamical scaling in Ising and vector spin glasses}},
  journal = {Phys. Rev. B},
  volume  = {72},
  pages   = {014462},
  year    = {2005},
  doi     = {10.1103/PhysRevB.72.014462}
}

@article{Alba2011,
  author  = {Alba, Vincenzo and Vicari, Ettore},
  title   = {{Temperature-disorder phase diagram of a three-dimensional gauge-glass model}},
  journal = {Phys. Rev. B},
  volume  = {83},
  pages   = {094203},
  year    = {2011},
  doi     = {10.1103/PhysRevB.83.094203}
}

@article{Kondor1989,
doi = {10.1088/0305-4470/22/5/005},
year = {1989},
month = {mar},
publisher = {},
volume = {22},
number = {5},
pages = {L163},
author = {I Kondor},
title = {On chaos in spin glasses},
journal = {J. Phys. A},
}

@article{Ritort1994,
  author  = {Ritort, F.},
  title   = {{Static chaos and scaling behavior in the spin-glass phase}},
  journal = {Phys. Rev. B},
  volume  = {50},
  number  = {10},
  pages   = {6844},
  year    = {1994},
  doi     = {10.1103/PhysRevB.50.6844}
}

@article{Billoire2003,
  author  = {Billoire, A. and Coluzzi, B.},
  title   = {{Magnetic field chaos in the Sherrington--Kirkpatrick model}},
  journal = {Phys. Rev. E},
  volume  = {67},
  pages   = {036108},
  year    = {2003},
  doi     = {10.1103/PhysRevE.67.036108}
}

@article{AguilarJanita2024,
  author  = {Aguilar-Janita, Miguel and Franz, Silvio and Martin-Mayor, Victor and Moreno-Gordo, Javier and Parisi, Giorgio and Ricci-Tersenghi, Federico and Ruiz-Lorenzo, Juan J.},
  title   = {{Small field chaos in spin glasses: Universal predictions from the ultrametric tree and comparison with numerical simulations}},
  journal = {Proc. Natl. Acad. Sci. USA},
  volume  = {121},
  number  = {40},
  pages   = {e2404973121},
  year    = {2024},
  doi     = {10.1073/pnas.2404973121}
}

@article{Fisher1986,
  title = {Ordered Phase of Short-Range {Ising} Spin-Glasses},
  author = {Fisher, Daniel S. and Huse, David A.},
  journal = {Phys. Rev. Lett.},
  volume = {56},
  issue = {15},
  pages = {1601},
  numpages = {0},
  year = {1986},
  month = {Apr},
  publisher = {American Physical Society},
  doi = {10.1103/PhysRevLett.56.1601},
}

@article{Fisher1988,
author = {Fisher, D S and Huse, D A},
journal = {Phys. Rev. B},
number = {1},
pages = {386},
title = {{Equilibrium behavior of the spin-glass ordered phase}},
volume = {38},
doi={10.1103/PhysRevB.38.386},
year = {1988}
}

@article{Bray1987,
  title = {Chaotic Nature of the Spin-Glass Phase},
  author = {Bray, A. J. and Moore, M. A.},
  journal = {Phys. Rev. Lett.},
  volume = {58},
  issue = {1},
  pages = {57},
  numpages = {0},
  year = {1987},
  month = {Jan},
  publisher = {American Physical Society},
  doi = {10.1103/PhysRevLett.58.57},
}

@article{Imry1975,
  author  = {Imry, Y. and Ma, S.-K.},
  title   = {{Random-field instability of the ordered state of continuous symmetry}},
  journal = {Phys. Rev. Lett.},
  volume  = {35},
  number  = {21},
  pages   = {1399},
  year    = {1975},
  doi     = {10.1103/PhysRevLett.35.1399}
}

@article{AizenmanWehr1989,
  author  = {Aizenman, Michael and Wehr, Jan},
  title   = {{Rounding of first-order phase transitions in systems with quenched disorder}},
  journal = {Phys. Rev. Lett.},
  volume  = {62},
  number  = {21},
  pages   = {2503},
  year    = {1989},
  doi     = {10.1103/PhysRevLett.62.2503}
}

@article{AizenmanWehr1990,
  author  = {Aizenman, Michael and Wehr, Jan},
  title   = {{Rounding effects of quenched randomness on first-order phase transitions}},
  journal = {Commun. Math. Phys.},
  volume  = {130},
  number  = {3},
  pages   = {489},
  year    = {1990},
  doi     = {10.1007/BF02096933}
}

@article{Dario2024,
  author  = {Dario, Paul and Harel, Matan and Peled, Ron},
  title   = {{Quantitative disorder effects in low-dimensional spin systems}},
  journal = {Commun. Math. Phys.},
  volume  = {405},
  pages   = {212},
  year    = {2024},
  doi     = {10.1007/s00220-024-05081-9}
}

@article{AizenmanHarelPeled2020,
  author  = {Aizenman, Michael and Harel, Matan and Peled, Ron},
  title   = {{Exponential decay of correlations in the 2D random field Ising model}},
  journal = {J. Stat. Phys.},
  volume  = {180},
  pages   = {304},
  year    = {2020},
  doi     = {10.1007/s10955-019-02401-5}
}

@article{DingXia2021,
  author  = {Ding, Jian and Xia, Jiaming},
  title   = {{Exponential decay of correlations in the two-dimensional random field Ising model}},
  journal = {Invent. Math.},
  volume  = {224},
  number  = {3},
  pages   = {999},
  year    = {2021},
  doi     = {10.1007/s00222-020-01024-y}
}

@article{Hartmann2001,
author = {Hartmann, A. K. and Young, A. P.},
doi = {10.1103/PhysRevB.64.180404},
journal = {Phys. Rev. B},
month = {oct},
number = {18},
pages = {180404(R)},
title = {{Lower critical dimension of Ising spin glasses}},
volume = {64},
year = {2001}
}

@article{Jorg2006,
  title = {Strong Universality and Algebraic Scaling in Two-Dimensional {Ising} Spin Glasses},
  author = {J\"org, T. and Lukic, J. and Marinari, E. and Martin, O. C.},
  journal = {Phys. Rev. Lett.},
  volume = {96},
  issue = {23},
  pages = {237205},
  numpages = {4},
  year = {2006},
  month = {Jun},
  publisher = {American Physical Society},
  doi = {10.1103/PhysRevLett.96.237205},
}

@article{Edwards1975,
doi = {10.1088/0305-4608/5/5/017},
year = {1975},
volume = {5},
pages = {965},
author = {S F Edwards and  P W Anderson},
title = {Theory of spin glasses},
journal = {J. Phys. F},
}

@article{Sherrington1975,
author = {Sherrington, D and Kirkpatrick, S},
doi = {10.1103/PhysRevLett.35.1792},
journal = {Phys. Rev. Lett.},
pages = {1792},
title = {{Solvable model of a spin-glass}},
volume = {35},
year = {1975}
}

@article{Parisi1979,
author = {Parisi, Giorgio},
title = {{Infinite number of order parameters for spin-glasses}},
journal = {Phys. Rev. Lett.},
volume = {43},
number = {23},
pages = {1754},
year = {1979},
doi = {10.1103/PhysRevLett.43.1754}
}

@article{Parisi1980a,
author = {Parisi, Giorgio},
title = {{The order parameter for spin glasses: a function on the interval 0-1}},
journal = {J. Phys. A},
volume = {13},
number = {3},
pages = {1101},
year = {1980},
doi = {10.1088/0305-4470/13/3/042}
}

@article{deAlmeida1978,
author = {de Almeida, J. R. L. and Thouless, D. J.},
title = {{Stability of the {Sherrington--Kirkpatrick} solution of a spin glass model}},
journal = {J. Phys. A},
volume = {11},
number = {5},
pages = {983},
year = {1978},
doi = {10.1088/0305-4470/11/5/028}
}

@book{Mezard1987,
  author = {M{\'e}zard, Marc and Parisi, Giorgio and Virasoro, Miguel Angel},
  title = {{Spin Glass Theory and Beyond}},
  publisher = {World Scientific},
  address = {Singapore},
  year = {1987}
}

@book{Charbonneau2023,
  title={Spin Glass Theory and Far Beyond: Replica Symmetry Breaking after 40 Years},
  editor={Charbonneau, Patrick and Marinari, Enzo and M\'ezard, Marc and Parisi, Giorgio and Ricci-Tersenghi, Federico and Sicuro, Gabriele and Zamponi, Francesco},
  year={2023},
  address = {Singapore},
  publisher={World Scientific}
}

@book{SteinNewman2013,
  author    = {Stein, Daniel L. and Newman, Charles M.},
  title     = {{Spin Glasses and Complexity}},
  publisher = {Princeton University Press},
  address   = {Princeton},
  year      = {2013}
}

@article{GuerraToninelli2002,
  author  = {Guerra, Francesco and Toninelli, Fabio Lucio},
  title   = {{The thermodynamic limit in mean field spin glass models}},
  journal = {Commun. Math. Phys.},
  volume  = {230},
  pages   = {71},
  year    = {2002},
  doi     = {10.1007/s00220-002-0699-y}
}

@article{Horiguchi1982,
doi = {10.1088/0305-4470/15/2/005},
year = {1982},
volume = {15},
number = {2},
pages = {L75},
author = {T Horiguchi and  T Morita},
title = {Existence of the ferromagnetic phase in a random-bond {Ising} model on the square lattice},
journal = {J. Phys. A}
}

@article{Nishimori1986,
author = {Nishimori, Hidetoshi},
journal = {J. Phys. Soc. Jpn.},
number = {10},
pages = {3305},
title = {{Geometry-induced phase transition in the $\pm J$ Ising model}},
volume = {55},
year = {1986},
doi={10.1143/JPSJ.55.3305}
}

@article{Kitatani1992,
author = {Kitatani, Hidetsugu},
doi = {10.1143/JPSJ.61.4049},
journal = {J. Phys. Soc. Jpn.},
number = {11},
pages = {4049},
title = {The Verticality of the Ferromagnetic-Spin Glass Phase Boundary of the {$\pm J$ Ising} Model in the {$p$-$T$} Plane},
volume = {61},
year = {1992}
}

@article{Nishimori1983,
author = {Nishimori, Hidetoshi and Stephen, Michael J},
journal = {Phys. Rev. B},
number = {9},
pages = {5644},
title = {{Gauge-invariant frustrated Potts spin-glass}},
volume = {27},
doi = {10.1103/PhysRevB.27.5644},
year = {1983}
}

@article{Garban2022,
  title={Continuous symmetry breaking along the {Nishimori} line},
  author={Garban, Christophe and Spencer, Thomas},
  journal={J. Math. Phys.},
  volume={63},
  number={9},
  year={2022},
  pages={093302},
  doi={10.1063/5.0087024}
}

@article{Nishimori2026pspin,
author = {Nishimori, Hidetoshi},
title = {{Reentrance and temperature chaos in the $p$-spin Ising spin glass}},
journal = {Phys. Rev. E},
volume = {114},
pages = {044127},
year = {2026},
doi = {10.1103/1xrx-rsxn}
}

@article{Ohzeki2009,
author = {Ohzeki, Masayuki and Nishimori, Hidetoshi},
doi = {10.1088/1751-8113/42/33/332001},
journal = {J. Phys. A},
number = {33},
pages = {332001},
title = {{Analytical evidence for the absence of spin glass transition on self-dual lattices}},
volume = {42},
year = {2009}
}

@article{Hoyos2011,
  title={Protecting clean critical points by local disorder correlations},
doi={10.1209/0295-5075/93/30004},
  author={Hoyos, Jos{\'e} A and Laflorencie, Nicolas and Vieira, Andr{\'e} P and Vojta, Thomas},
  journal={Europhys. Lett.},
  volume={93},
  number={3},
  pages={30004},
  year={2011}
}

@article{Bonzom2013,
  title={Universality in {$p$}-spin glasses with correlated disorder},
  doi={10.1088/1742-5468/2013/02/L02003},
  author={Bonzom, Valentin and Gurau, Razvan and Smerlak, Matteo},
  journal={J. Stat. Mech.},
  volume={2013},
  number={02},
  pages={L02003},
  year={2013}
}

@article{Cavaliere2019,
doi = {10.1088/1751-8121/ab10f9},
year = {2019},
volume = {52},
number = {17},
pages = {174002},
author = {Angelo Giorgio Cavaliere and Andrea Pelissetto},
title = {Disordered {Ising} model with correlated frustration},
journal = {J. Phys. A},
}

@article{Munster2021,
  title = {Ordering behavior of the two-dimensional {Ising} spin glass with long-range correlated disorder},
  author = {M\"unster, L. and Norrenbrock, C. and Hartmann, A. K. and Young, A. P.},
  journal = {Phys. Rev. E},
  volume = {103},
  pages = {042117},
  year = {2021},
  doi = {10.1103/PhysRevE.103.042117}
}

@article{Banavar1987,
  title = {Chaos in spin glasses: A renormalization-group study},
  author = {Banavar, Jayanth R. and Bray, Alan J.},
  journal = {Phys. Rev. B},
  volume = {35},
  pages = {8888},
  year = {1987},
  doi = {10.1103/PhysRevB.35.8888}
}

@article{Rizzo2003,
author = {Rizzo, T. and Crisanti, A.},
doi = {10.1103/PhysRevLett.90.137201},
journal = {Phys. Rev. Lett.},
number = {13},
pages = {137201},
title = {{Chaos in temperature in the Sherrington-Kirkpatrick model}},
volume = {90},
year = {2003}
}

@article{Katzgraber2007b,
  title = {Temperature and Disorder Chaos in Three-Dimensional {Ising} Spin Glasses},
  author = {Katzgraber, Helmut G. and Krz\k{a}ka\l{}a, Florent},
  journal = {Phys. Rev. Lett.},
  volume = {98},
  pages = {017201},
  year = {2007},
  doi = {10.1103/PhysRevLett.98.017201}
}

@article{Fernandez2013,
doi = {10.1209/0295-5075/103/67003},
year = {2013},
volume = {103},
number = {6},
pages = {67003},
author = {L. A. Fernandez and V. Martin-Mayor and G. Parisi and B. Seoane},
title = {Temperature chaos in {3D} {Ising} spin glasses is driven by rare events},
journal = {Europhys. Lett.},
}

@article{BaityJesi2021,
author = {Baity-Jesi, Marco and Calore, Enrico and Cruz, Andr{\'{e}}s and Fernandez, Luis Antonio and Gil-Narvion, Jos{\'{e}} Miguel and {Gonzalez-Adalid Pemartin}, Isidoro and Gordillo-Guerrero, Antonio and I{\~{n}}iguez, David and Maiorano, Andrea and Marinari, Enzo and Martin-Mayor, V{\'{i}}ctor and Moreno-Gordo, Javier and Mu{\~{n}}oz-Sudupe, Antonio and Navarro, Denis and Paga, Ilaria and Parisi, Giorgio and Perez-Gaviro, Sergio and Ricci-Tersenghi, Federico and Ruiz-Lorenzo, Juan Jes{\'{u}}s and Schifano, Sebastiano Fabio and Seoane, Beatriz and Tarancon, Alfonso and Tripiccione, Raffaele and Yllanes, David},
doi = {10.1038/s42005-021-00565-9},
journal = {Commun. Phys.},
number = {1},
pages = {74},
title = {{Temperature chaos is present in off-equilibrium spin-glass dynamics}},
volume = {4},
year = {2021}
}

@article{Chen2013,
author = {Chen, Wei-Kuo and Panchenko, Dmitry},
title = {{An approach to chaos in some mixed $p$-spin models}},
journal = {Probab. Theory Relat. Fields},
volume = {157},
number = {1-2},
pages = {389},
year = {2013},
doi = {10.1007/s00440-012-0460-1}
}

@article{Panchenko2016,
author = {Panchenko, Dmitry},
title = {{Chaos in temperature in generic $2p$-spin models}},
journal = {Commun. Math. Phys.},
volume = {346},
number = {2},
pages = {703},
year = {2016},
doi = {10.1007/s00220-016-2585-z}
}

@article{Nobre2001,
  author    = {Fernando D. Nobre},
  title     = {Phase diagram of the two-dimensional {$\pm J$ Ising} spin glass},
  journal   = {Phys. Rev. E},
  volume    = {64},
  number    = {4},
  pages     = {046108},
  year      = {2001},
  doi       = {10.1103/PhysRevE.64.046108}
}

@article{Hasenbusch2007,
  author    = {Martin Hasenbusch and Francesco {Parisen Toldin} and Andrea Pelissetto and Ettore Vicari},
  title     = {Critical behavior of the three-dimensional {$\pm J$ Ising} model at the paramagnetic-ferromagnetic transition line},
  journal   = {Phys. Rev. B},
  volume    = {76},
  number    = {9},
  pages     = {094402},
  year      = {2007},
  doi       = {10.1103/PhysRevB.76.094402}
}

@article{Ceccarelli2011,
author = {Ceccarelli, Giacomo and Pelissetto, Andrea and Vicari, Ettore},
doi = {10.1103/PhysRevB.84.134202},
journal = {Phys. Rev. B},
number = {13},
pages = {134202},
title = {{Ferromagnetic-glassy transitions in three-dimensional Ising spin glasses}},
volume = {84},
year = {2011}
}

@article{FisherMPA1989,
  author  = {Fisher, Matthew P. A.},
  title   = {{Vortex-glass superconductivity: A possible new phase in bulk high-$T_c$ oxides}},
  journal = {Phys. Rev. Lett.},
  volume  = {62},
  pages   = {1415},
  year    = {1989},
  doi     = {10.1103/PhysRevLett.62.1415}
}

@article{HuseSeung1990,
  author  = {Huse, David A. and Seung, H. Sebastian},
  title   = {{Possible vortex-glass transition in a model random superconductor}},
  journal = {Phys. Rev. B},
  volume  = {42},
  pages   = {1059(R)},
  year    = {1990},
  doi     = {10.1103/PhysRevB.42.1059}
}

@article{KosterlitzAkino1998,
  author  = {Kosterlitz, J. M. and Akino, N.},
  title   = {{Numerical study of order in a gauge glass model}},
  journal = {Phys. Rev. Lett.},
  volume  = {81},
  pages   = {4672},
  year    = {1998},
  doi     = {10.1103/PhysRevLett.81.4672}
}

@article{YoungKatzgraber2004,
  author  = {Young, A. P. and Katzgraber, Helmut G.},
  title   = {{Absence of an Almeida-Thouless line in three-dimensional spin glasses}},
  journal = {Phys. Rev. Lett.},
  volume  = {93},
  pages   = {207203},
  year    = {2004},
  doi     = {10.1103/PhysRevLett.93.207203}
}

@article{MooreBray2011,
  author  = {Moore, M. A. and Bray, A. J.},
  title   = {{Disappearance of the de Almeida-Thouless line in six dimensions}},
  journal = {Phys. Rev. B},
  volume  = {83},
  pages   = {224408},
  year    = {2011},
  doi     = {10.1103/PhysRevB.83.224408}
}

@article{Janus2012,
  author  = {{Alvarez Ba{\~n}os}, Raquel and others},
  title   = {{Thermodynamic glass transition in a spin glass without time-reversal symmetry}},
  journal = {Proc. Natl. Acad. Sci. USA},
  volume  = {109},
  pages   = {6452},
  year    = {2012},
  doi     = {10.1073/pnas.1203295109}
}

@article{MerminWagner1966,
  author  = {Mermin, N. D. and Wagner, H.},
  title   = {{Absence of ferromagnetism or antiferromagnetism in one- or two-dimensional isotropic Heisenberg models}},
  journal = {Phys. Rev. Lett.},
  volume  = {17},
  pages   = {1133},
  year    = {1966},
  doi     = {10.1103/PhysRevLett.17.1133}
}

@article{AGL2012,
  author  = {Aizenman, Michael and Greenblatt, Rafael L. and Lebowitz, Joel L.},
  title   = {{Proof of rounding by quenched disorder of first order transitions in low-dimensional quantum systems}},
  journal = {J. Math. Phys.},
  volume  = {53},
  pages   = {023301},
  year    = {2012},
  doi     = {10.1063/1.3679069}
}

@article{Chatterjee2018,
  author  = {Chatterjee, Sourav},
  title   = {{On the decay of correlations in the random field Ising model}},
  journal = {Commun. Math. Phys.},
  volume  = {362},
  pages   = {253},
  year    = {2018},
  doi     = {10.1007/s00220-018-3085-0}
}

@article{AizenmanPeled2019,
  author  = {Aizenman, Michael and Peled, Ron},
  title   = {{A power-law upper bound on the correlations in the 2D random field Ising model}},
  journal = {Commun. Math. Phys.},
  volume  = {372},
  pages   = {865},
  year    = {2019},
  doi     = {10.1007/s00220-019-03450-3}
}

@article{Imbrie1985,
  author  = {Imbrie, John Z.},
  title   = {{The ground state of the three-dimensional random-field Ising model}},
  journal = {Commun. Math. Phys.},
  volume  = {98},
  pages   = {145},
  year    = {1985},
  doi     = {10.1007/BF01220505}
}

@article{BricmontKupiainen1988,
  author  = {Bricmont, J. and Kupiainen, A.},
  title   = {{Phase transition in the 3d random field Ising model}},
  journal = {Commun. Math. Phys.},
  volume  = {116},
  pages   = {539},
  year    = {1988},
  doi     = {10.1007/BF01224901}
}

@article{DingZhuang2024,
  author  = {Ding, Jian and Zhuang, Zijie},
  title   = {{Long range order for random field Ising and Potts models}},
  journal = {Commun. Pure Appl. Math.},
  volume  = {77},
  pages   = {37},
  year    = {2024},
  doi     = {10.1002/cpa.22127}
}

@article{GingrasHuse1996,
  author  = {Gingras, Michel J. P. and Huse, David A.},
  title   = {{Topological defects in the random-field XY model and the pinned vortex lattice to vortex glass transition in type-II superconductors}},
  journal = {Phys. Rev. B},
  volume  = {53},
  pages   = {15193},
  year    = {1996},
  doi     = {10.1103/PhysRevB.53.15193}
}

@article{Feldman2000,
  author  = {Feldman, D. E.},
  title   = {{Nonexistence of quasi-long-range order in strongly disordered vortex glasses: A rigorous proof}},
  journal = {Phys. Rev. B},
  volume  = {62},
  pages   = {5364},
  year    = {2000},
  doi     = {10.1103/PhysRevB.62.5364}
}

@article{TissierTarjus2006,
  author  = {Tissier, Matthieu and Tarjus, Gilles},
  title   = {{Unified picture of ferromagnetism, quasi-long-range order, and criticality in random-field models}},
  journal = {Phys. Rev. Lett.},
  volume  = {96},
  pages   = {087202},
  year    = {2006},
  doi     = {10.1103/PhysRevLett.96.087202}
}

@article{Oshima2026,
  author  = {Oshima, Takumi and Arai, Yamato and Hukushima, Koji},
  title   = {{Phase transitions in a modified Ising spin glass model: A tensor-network-based sampling approach}},
  journal = {arXiv:2603.18486},
  url     = {https://arxiv.org/abs/2603.18486},
  year    = {2026}
}

\end{document}